\documentclass[12pt]{iopart}

\usepackage{iopams}  
\usepackage{setstack}
\usepackage{graphicx}
\usepackage{bm}
\usepackage[utf8]{inputenc} 
\usepackage{xcolor}
\usepackage[colorlinks, citecolor=blue, urlcolor=blue, linkcolor=black,breaklinks]{hyperref}
\usepackage[normalem]{ulem}

\newcommand{\mo}{moir\'{e}}
\newcommand{\Mo}{Moir\'{e}}
\usepackage{soul}
\usepackage{cite}
\begin{document}
\setstcolor{red}
\title{Optical properties and plasmons in {\mo} structures}

\author{Xueheng Kuang}
\address{Yangtze Delta Industrial Innovation Center of Quantum Science and Technology, Suzhou 215000, China}
\author{Pierre A. Pantale\'on Peralta}
\address{Instituto Madrile\~no de Estudios Avanzados, IMDEA Nanociencia, Calle Faraday 9, 28049, Madrid, Spain}
\author{Jose Angel Silva-Guill\'en}
\address{Instituto Madrile\~no de Estudios Avanzados, IMDEA Nanociencia, Calle Faraday 9, 28049, Madrid, Spain}
\author{Shengjun Yuan}
\address{Key Laboratory of Artificial Micro- and Nano-structures of the Ministry of Education and School of Physics and Technology, Wuhan University, Wuhan 430072, China}
\address{Wuhan Institute of Quantum Technology, Wuhan 430206, China}
\author{Francisco Guinea}
\address{Instituto Madrile\~no de Estudios Avanzados, IMDEA Nanociencia, Calle Faraday 9, 28049, Madrid, Spain}
\address{Donostia International Physics Center, Paseo Manuel de Lardiz\'abal 4, 20018, San Sebasti\'an, Spain}
\author{Zhen Zhan}
\address{Instituto Madrile\~no de Estudios Avanzados, IMDEA Nanociencia, Calle Faraday 9, 28049, Madrid, Spain}
\ead{zhenzhanh@gmail.com}
\vspace{10pt}
\begin{indented}
	\item[]November 2023
\end{indented}

\newpage

\begin{abstract}
The discoveries of numerous exciting phenomena in twisted bilayer graphene (TBG) are stimulating significant investigations on {\mo} structures that possess a tunable {\mo} potential. 
Optical response can provide insights into the electronic structures and transport phenomena of non-twisted and twisted {\mo} structures. 
In this article, we review both experimental and theoretical studies of optical properties such as optical conductivity, dielectric function, non-linear optical response, and plasmons in {\mo} structures composed of graphene, hexagonal boron nitride (hBN), and/or transition metal dichalcogenides (TMDCs). Firstly, a comprehensive introduction to the widely employed methodology on optical properties is presented. After, {\mo} potential induced optical conductivity and plasmons in non-twisted  structures are reviewed, such as single layer graphene-hBN, bilayer graphene-hBN and graphene-metal {\mo} heterostructures. Next, recent investigations of twist-angle dependent optical response and plasmons are addressed in twisted {\mo} structures. Additionally, we discuss how optical properties and plasmons could contribute to the understanding of  the many-body effects and superconductivity observed in {\mo} structures. 
\end{abstract}

%
%
%
%
%
\newpage


\section{Introduction} 
Two-dimensional (2D) materials, such as graphene\cite{novoselov2004electric}, hexagonal boron nitride (hBN)\cite{Lin2010soluble,Weng2016functionalized}, transition metal dichalcogenides (TMDCs)\cite{Chhowalla2013thechemis,Manzeli20172D} and many others\cite{Li2014black,ji2016two}, have been widely investigated over the past two decades \cite{roldan2017theory}. 
The fact that the isolated atomic planes could be stacked layer by layer via weak van der Waals interactions, forming moir\'e structures, provides an avenue to realize functionalities distinct from the components of the moir\'e systems \cite{Geim2013van}. 
Interestingly, heterostructures of two materials with different lattice constants form large-scale structures that display a {\mo} pattern, which we will refer to as non-twisted {\mo} structures. For example, by depositing graphene on top of hBN, if they are aligned, we would obtain a graphene-hBN {\mo} structure with a {\mo} length of around 14 nm \cite{dean2012graphene}.
{\Mo} structures can also be obtained by rotating two layers with a relative twist angle, which are commonly known as twisted {\mo} structures \cite{andrei2020graphene,andrei2021marvels}.
Twisted {\mo} structures have attracted tremendous attentions due to the emergence of a rich phase diagram of correlated states. For instance, in twisted bilayer graphene (TBG), when two graphene layers are rotated by an angle of 1.05$^\circ$, known as the magic angle, flat bands appear at the charge neutrality point\cite{SuarezMorell2010flat,Bistritzer2011moire}. As a consequence, exotic phenomena are observed, ranging from unconventional superconductivity\cite{cao2018unconventional,yankowitz2019tuning,Oh2021evidence} to correlated insulator phases\cite{cao2018correlated,choi2019electronic,Xie2019spectroscopic,Kerelski2019maximized,Liu2021tuning}, topological Chern insulators\cite{Xie2021fractional,Choi2021correlation}, ferromagnetism\cite{Pons2020flatband,Bhowmik2022spinorbit,Lin2022spinorbit}, anomalous Hall effects\cite{li2020quantum,Tseng2022anomalous,Bhowmik2022spinorbit} and non-linear Hall effects\cite{Duan2022giant}.
These findings have also motivated further search of electronic flat bands in other twisted {\mo} structures like twisted trilayer graphene (TTG)\cite{zhu2020twisted,Park2020gatetunable,wu2021lattice}, twisted double bilayer graphene (TDBG) \cite{chebrolu2019flat,Rickhaus2019gap,Haddadi2020moire,Rickhaus2021correlated}, twisted bilayer TMDCs\cite{naik2018ultraflatbands,Venkateswarlu2020electronic,Zhan2020tunability,vitale2021flat,Shabani2021deepmoire,Devakul2021magic,Zhang2021electronic}, twisted bilayer hBN\cite{Xian2019multiflat,zhao2020formation,walet2021flat,Woods2021charge}, and so on. In fact, correlated insulators and tunable superconductivity have also been found in twisted multilayer graphene systems\cite{Cao2020tunable,shen2020correlated,park2021tunable,He2021symmetry,kim2022evidence,hao2021electric,chen2021electrically,Park2022robust,Zhang2022promotion,Burg2022emergence}.  Emergence of flat bands can also induce orbital ferromagnetism and correlated insulator states in non-twisted {\mo} structures such as aligned trilayer ABC graphene-hBN \cite{zhu2018inter,Chen2019evidence,Chittari2019gatetunable,chen2020tunable}, and hBN-graphene-hBN heterostructures\cite{sun2021correlated}. 
On the other hand, twisted bilayer TMDCs can also be exceptional frameworks for investigating many-body insulators\cite{Wang2020correlated,Zhang2020flat}, Hubbard physics\cite{Zang2021hartree,Xu2022atunable} and the quantum anomalous Hall effect\cite{cai2023signatures,zeng2023thermodynamic,park2023observation,xu2023observation}. More importantly, large-scale {\mo} structures also enables exciting photonic and optoelectronic properties, like {\mo} excitons\cite{Andersen2021excitons} and polaritons as, reviewed in \cite{Du2023moire}.

In the search for downscaling technological devices, optical properties, such as optical conductivity\cite{mak2008measurement,mak2012optical,Rukelj2016optical}, dielectric function\cite{Johari2011tunable,li2014measurement,laturia2018dielectric,Scholz2013plasmons}, etc. have been extensively explored in 2D materials. 
The unique electronic structure of graphene has motivated research focused on the fields of non-linear plasmon response\cite{cox2016quantum}, plasmon-polariton\cite{koppens2011graphene,xiao2016graphene}, and other plasmonics\cite{gonccalves2016introduction}. From the perspective of applications, graphene stands out to be a very promising candidate for terahertz to mid-infrared applications\cite{reserbat2021quantum}, such as modulators, polarizers, mid-infrared photodetectors or mid-infrared vibrational spectroscopy \cite{Low2014graphene,vakil2011transformation}.
Furthermore, to make significant advances in the confinement of light, plasmons have been extensively studied in graphene and TMDCs and have been shown to have potential applications for the development of new state-of-the-art optical devices\cite{Low2014graphene,stauber2014plasmonics,Wang2015plasmon,groenewald2016valley,li2017plasmonics,Torbatian2017plasmon}.
From a fundamental point of view, the optical and plasmon properties of 2D materials are extremely sensitive to their band structure. Thus, the optical properties could shed light on electronic structure\cite{sunku2020nano}, which could serve to further investigate exotic quantum phases. Since {\mo} structures can exhibit distinct electronic properties from their building block counterparts, {\mo} structures such as TBG can lead to numerous exciting optical phenomena and could be used in future generations of optoelectronic devices\cite{catarina2019twisted,deng2020strong}.
Here, we present an overview of the recent progress in the matter of emerging linear optical response, plasmons and their relations to other interesting properties in both non-twisted and twisted {\mo} structures.

This review is organized as follows: Section 2 introduces the common experimental and theoretical methods for investigating optical properties and plasmons. Optical properties and plasmons are described for non-twisted {\mo} structures, and for twisted {\mo} structures in Sections 3 and 4, respectively. The relation to other properties such as many-body effects and superconductivity, and a brief description of the non-linear optical response are discussed in Section 5. In Section 6 we give a final summary and future perspectives.

\section{Methodology} 
This section is devoted to the introduction of the experimental techniques and theoretical framework that are needed to study the linear optical response of {\mo} structures. 

\subsection{Experimental techniques}
In general, 2D materials and 2D material-based {\mo} structures are expected to show strong light-matter interaction and enriched photoresponses. For 2D systems, the response to an applied electromagnetic field can be mainly characterized by the optical conductivity, $\sigma(\omega)$. 
Since the optical conductivity is uniquely determined by the band structure, it is a powerful tool to understand the electronic properties of  materials.
Experimentally, infrared spectroscopy (IR) is a widely used technique to measure the optical conductivity of a material \cite{Low2014graphene}. 
Another promising technique is the scattering-type scanning near-field optical microscope (s-SNOM)\cite{aizpurua2008substrate}, which could provide the propagation of the surface plasmons by measuring the scattering amplitude $S_{opt}(x)$. 
The advantage of the s-SNOM is that, from the plasmon dispersion, it is possible to extract the optical conductivity $\sigma(\omega)$ and loss function $S(\mathbf{q},\omega)$, enabling experimental access to both the electronic band structure and electron-hole excitations of the systems \cite{Ni2015plasmons,Hesp2021observation}. 
Electron energy-loss spectroscopy (EELS) has also advanced in recent decades to provide the structural and optical characterization of materials by correlating the acquired infrared-to-ultraviolet spectral data with morphological and structural information derived from secondary electron images (in scanning electron microscope (SEM)) or the high-angle annular dark-field signal (in transmission electron microscope (TEM)) \cite{polman2019electron}. 
The loss function $S(\mathbf{q},\omega)$ can be extracted from the EELS spectra. 
Currently, the measured results are a collection of signals from large area samples, which are unable to provide the local structure of the moir\'{e} system and could  be influenced by  extrinsic effects, for example, twist angle inhomogeneities or strain present in the sample. 
Thanks to the development of new techniques, such as 4D scanning TEM spectroscopy \cite{yan2021single} and near-field scanning \cite{Sunku2018photonic}, it is possible to locally probe the optoelectronic properties and could be used to study the properties of {\mo} structures.  

\subsection{Theoretical methods}
Concerning the theoretical approach, 
 through linear response theory, we can obtain optical properties by calculating different response functions. 
For example, the optical conductivity $\sigma(\omega)$, and the polarization function $\Pi_0(\mathbf{q},\omega)$ are evaluated from the current-current and non-interacting density-density response functions, respectively.
Other optical properties such as the dielectric function $\varepsilon(\mathbf{q},\omega)$, loss function $S(\mathbf{q},\omega)$ and the optical absorption coefficient can be extracted from those quantities, as discussed below. 

\subsubsection{Linear optical response}
The optical conductivity can be derived using the Kubo formula\cite{kubo1957statistical} and can be written as the Kubo-Greenwood equation\cite{Calderin2017kubo}
\begin{eqnarray}
	\fl \sigma _{\alpha1 \alpha2}\left( \omega \right) =\frac{g_si}{ (2\pi )^D}\int_{BZ}{d^D}k\sum_{l,l^{\mathrm{'}}}{\frac{n_F\left( E_{\boldsymbol{k}l^{\mathrm{'}}} \right) -n_F\left( E_{\boldsymbol{k}l} \right)}{E_{\boldsymbol{k}l}-E_{\boldsymbol{k}l^{\mathrm{'}}}}}{\frac{ \left< \boldsymbol{k}l^{\mathrm{'}}\left| J_{\alpha1} \right|\boldsymbol{k}l \right> \left< \boldsymbol{k}l\left| J_{\alpha2} \right|\boldsymbol{k}l^{\mathrm{'}} \right> }{E_{\boldsymbol{k}l^{\mathrm{'}}}-E_{\boldsymbol{k}l}+\hbar \omega +i\delta}}, \label{optic}
\end{eqnarray}
where $g_s$ is the spin degeneracy, $D$ is the dimension of {\mo} structures and is typically set to 2 for 2D materials. 
$J_{\alpha1}$ and $J_{\alpha2}$ are current operators along the $\alpha1$ and $\alpha2$ directions, respectively. $n_F$ is the Fermi-Dirac distribution. Eigenvalues, $E_{\boldsymbol{k}l}$, and eigenstates, $|\boldsymbol{k}l \rangle$,  with band index $l$ and momentum $\boldsymbol{k}$, are needed to describe optical band transitions between $l$ and $l^{\mathrm{'}}$ bands.
The integration runs over the whole Brillouin zone (BZ).

By combining the Kubo formula with the tight-binding propagation method (TBPM), the optical conductivity (omitting the Drude contribution at $\omega=0$) could be expressed as \cite{Yuan2010modeling,Li2023tbplas}
\begin{eqnarray}
	\fl \sigma_{\alpha1 \alpha2}\left( \omega \right) =\underset{\varepsilon \rightarrow 0^+}{\mathbf{lim}}\frac{e^{-\beta \hbar \omega}-1}{\hbar \omega \Omega}\int_0^{\infty}{e^{-\varepsilon t}}\left( \sin\omega t-i\cos\omega t \right) \nonumber \\
	\times 2\mathrm{Im}\left\{\langle \varphi \left| n_F\left( H \right) e^{iHt}J_{\alpha1}e^{-iHt}[1-n_F\left( H \right) ]J_{\alpha2} \right|\varphi \rangle\right\} dt, \label{tbpm_optic}
\end{eqnarray}
where $\Omega$ is the area of the system, $\beta=1/(k_BT)$ being $k_B$ the Boltzmann constant, $H$ is the Hamiltonian, and $|\varphi\rangle$ is the
 initial state of the system, which is a random superposition of all basis states 
\begin{equation}
	|\varphi\rangle = \sum_{i}a_i|i\rangle, \label{random}
\end{equation}
where ${|i\rangle}$ are all basis states in real space and $a_i$ are random complex numbers normalized as $\sum_{i}|a_i|^2 = 1$. 

The calculation of equation (\ref{tbpm_optic}) scales linearly with the number of states $N$ of the system in real space. 
In contrast, the scaling would be $O(N^3)$ if we were to solve equation (\ref{optic}) using the exact diagonalization method to obtain the eigenstates and eigenvalues of the system.
Therefore, calculating the optical conductivity using equation (\ref{tbpm_optic}) has advantages when dealing with non-periodic moir\'{e} structures, such as 30$^\circ$ dodecagonal graphene quasicrystal \cite{yu2019dodecagonal,ahn2018dirac,yao2018quasicrystalline,pezzini202030,moon2019quasicrystalline} and large-scale periodic moir\'{e} structures \cite{Li2023tbplas}.
Interestingly, disorder effects on optical conductivity can be also easily be considered with this real-space method~\cite{yuan2011optical}.
It is important to note that there are similar real-space methods with $O(N)$ time scaling that do not require time propagation to calculate transport conductivity $\sigma_{\alpha \beta}\left( \omega=0 \right)$ \cite{roche1999quantum,fan2021linear}.

The optical conductivity corresponds to the optical absorption spectrum that can be extracted from raw data of IR using multilayer Kramers-Kronig analysis program\cite{kuzmenko2005kramers}, and is related to the transmission of incident light perpendicular to the system, which is given by\cite{moon2013optical}
\begin{equation}
	T=|1 + \frac{2\pi}{c}\sigma(\omega)|^{-2}\approx1-\frac{4\pi}{c}\mathrm{Re}\left\{\sigma(\omega)\right\}.\label{Trans}
\end{equation}
Absorbance at normal incidence could be expressed as\cite{mak2012optical} 
\begin{equation}
	A = \frac{4\pi}{c}\mathrm{Re} \left\{\sigma(\omega)\right\}, \label{absorb}
\end{equation}
where $\mathrm{Re} \left\{\sigma(\omega)\right\}$ is the real part of optical conductivity.

\subsubsection{Polarization function}
The polarization function, $\Pi_0$, also known as charge susceptibility or non-interacting density-density response function, describes the charge fluctuation or single-particle transitions. 
Therefore, it is imperative to further describe collective excitations and screening in materials. 
For small-scale systems, such as large-angle twisted moir\'{e} structures whose eigenstates and eigenvalues can be obtained by diagonalization of the Hamiltonian, the polarization function can be solved by using the Lindhard function\cite{mahan2000many,giuliani2008quantum,coleman2015introduction}
\begin{eqnarray}
	\fl \Pi_0(\mathbf{q}, \omega) = &\frac{g_s}{(2\pi)^2}\int_\mathrm{BZ}d^2\mathbf{k}\sum_{l,l'}\frac{n_\mathrm{F}(E_{\mathbf{k'} l'}) - n_\mathrm{F}(E_{\mathbf{k}l})}
	{E_{\mathbf{k'} l'} - E_{\mathbf{k}l}-\omega-\mathrm{i}\delta}
	\times |\langle \mathbf{k'} l'|\mathrm e^{\mathrm{i}\mathbf{q\cdot r}}|\mathbf{k}l \rangle |^2,\label{lindhard}
\end{eqnarray}
where $\mathbf{k'}$=$\mathbf{k}$+$\mathbf{q}$, $\delta \rightarrow 0^+$. Generally, the integral is taken over the whole BZ, same as in equation (\ref{optic}). Note here that we named the polarization function without many-body effects and local field effects as $\Pi_0(\mathbf{q}, \omega)$.  

Combining the TBPM with the Kubo formula, the polarization function can also be described as \cite{Li2023tbplas,yuan2011excitation}
\begin{eqnarray}
	\fl \Pi_0(\mathbf{q},\omega)=& -\frac{2}{\Omega}\int_{0}^{\infty}\mathrm dt\; e^{i\omega t}\mathrm{Im}\langle \varphi| n_F(H)e^{iHt}
	\rho(\mathbf{q})e^{-iHt}[1-n_F(H)]\rho(-\mathbf{q})|\varphi\rangle, \label{kubo_dyn} 
\end{eqnarray} 
in which $\rho(\mathbf{q})=\sum_{i}c_i^{\dagger}c_i$exp$(i\mathbf{q}\cdot\mathbf{r}_i)$ is the density operator, $\mathbf{r}_i$ is the position of the $i$th orbital and $\Omega$ is the area of a {\mo} structure system, $|\varphi\rangle$ has the same form as equation (\ref{random}). 
Equation (\ref{kubo_dyn}) is equivalent to the Lindhard function (equation 
(\ref{lindhard})), which has been widely used in the study of single-layer graphene and twisted bilayer graphene \cite{Kuang2021collective}. 
The Lindhard function has the advantage that it can be used to study specific attributions of band transitions, such as the intraband and interband contributions to the polarization function in moir\'{e} structures, while this information cannot be extracted from equation (\ref{kubo_dyn}). 
However, when the full-band contribution to the polarization function is required to investigate screening effects in moir\'{e} structures, equation (\ref{kubo_dyn}) will be a reliable choice with lower computational complexity in comparison to the calculation of the Lindhard funtion (\ref{lindhard}) since it includes all possible electronic excitations in moir\'{e} structures. More details related to the TBPM and equation (\ref{kubo_dyn}) are discussed in \cite{Li2023tbplas}.

\subsubsection{Dielectric function and plasmons}

The dielectric function can be derived from the polarization function, $\Pi_0(\mathbf{q}, \omega)$, or the optical conductivity $\sigma(\omega)$. In the random phase approximation (RPA), the dielectric function relates directly to $\Pi_0(\mathbf{q},\omega)$ as
\begin{equation}
	\varepsilon(\mathbf{q}, \omega) = 1- V(q)\Pi_0(\mathbf{q}, \omega),\label{eps_dyn}
\end{equation}
where $V(q)$ is the Fourier component of the Coulomb interaction. For example, the pure two-dimensional Coulomb interaction is $V(q)=2\pi e^2/{(\varepsilon_\mathrm{B}}{q})$ with $\varepsilon_\mathrm{B}$ the background dielectric constant. Specifically, in the long wavelength limit $\mathbf{q} \rightarrow 0$, the RPA dielectric function is linked to $\sigma(\omega)$ as \cite{Stauber2013Optical}
\begin{equation}
\varepsilon\left(\mathbf{q},\omega\right)=1+\frac{iq^2V(q)}{\omega}\sigma\left(\omega\right). \label{eps_optic}
\end{equation}
Note that the accuracy of equation (\ref{eps_optic}) is lower when $q$ gets larger due to the local approximation used in the optical conductivity, but equation (\ref{eps_dyn}) is valid even for large $q$ because the polarization function is not a local property and dependent on $q$, whereas the optical conductivity $\sigma(\omega)$ is independent on $q$. The internal electronic screening potential can be given by the dielectric function as \cite{Maier2007plasmonics}
\begin{equation}
	V_{scr}(\mathbf{q}, \omega) = \frac{V(q)}{\varepsilon(\mathbf{q}, \omega)}.\label{V_scr}
\end{equation} 
A plasmon mode with momentum $\mathbf{q}$ and frequency $\omega_p$ can be obtained from the dielectric function with $\varepsilon(\mathbf{q}, \omega) = 0$ \cite{Maier2007plasmonics}.
The plasmon mode can be also measured from the electron energy loss function
\begin{equation}
	S(\mathbf{q}, \omega) = -\mathrm{Im}\frac{1}{\varepsilon(\mathbf{q}, \omega)}, \label{loss}
\end{equation}
with a sharp pole when $\omega = \omega_p$. 
The loss function is closely related to results obtained with EELS in experiments since the peaks in the data are related to the energy of the plasmon modes. Besides the experimental quantities of $\sigma_{\alpha \beta}( \omega)$ and $S(\mathbf{q}, \omega)$ that can be directly reproduced by the numerical calculations, these optical quantities could shed light on calculated electronic structures, for example, the bandwidth, band gap and Fermi velocity and so on, which is a good starting point to further understand exotic quantum phases.

Here, it should be noted that the limitation of RPA when the charge density is small and dimension of {\mo} structures is low, since vertex corrections could not be safely ignored in the dielectric function calculation\cite{morawetz2018conditions}. 
Based on many-body perturbation theory, the effects of vertex corrections on the dielectric function can be evaluated by taking into account the interaction between two non-interacting Green functions\cite{kotov2008electron,sabio2008f,abedinpour2011drude}, which are employed to derive the non-interacting density-density response function equation (\ref{lindhard}). Actually, previous studies have shown that vertex corrections has an impact on polarization function and plasmons in both doped and undoped graphene\cite{sabio2008f,gangadharaiah2008charge,abedinpour2011drude}. Beyond RPA, the exchange-correlation (EX) effects also play a role in affecting the dielectric function and plasmons\cite{inaoka2002two}. But, the EX effects can be incorporated into the RPA scheme using local-field corrections, which could be formulated by a dielectric function with a local-field factor that cannot be determined by a self-consistent calculation\cite{singwi1968electron,jonson1976electron,neilson1991dynamical,march1995collective}.

\subsubsection{Local field effects}
When confronted with inhomogeneous electron systems, it is crucial to consider the local field effects (LFE)\cite{adler1962quantum,wiser1963dielectric}, via an umklapp process in the analysis of optical properties and plasmons~\cite{louie1975local,van1972calculation,sturm1978local}. The local field effects become stronger as the momentum transfer $q$ increases, since then the wavelength of the excitation becomes smaller and one has to take the inhomogeneities of the electronic system under consideration. For example, for a {\mo} structure with large {\mo} length and/or large wavenumber $q$, when the investigated $q$ becomes comparable to the length of the first reciprocal {\mo} lattice vector, the LFE could significantly change the plasmon properties. The polarization function of including the LFE is given by 
\begin{eqnarray}
	\fl \Pi_{\mathbf{G}, \mathbf{G^\prime}}(\mathbf{q}, \omega) = \frac{g_s}{(2\pi)^2}\int_\mathrm{BZ}d^2\mathbf{k}\sum_{l,l'}\frac{n_\mathrm{F}(E_{\mathbf{k'} l'}) - n_\mathrm{F}(E_{\mathbf{k}l})}{E_{\mathbf{k'} l'} - E_{\mathbf{k}l}-\omega-\mathrm{i}\delta}\nonumber\\
	\times \langle \mathbf{k} l|\mathrm e^{-\mathrm{i}\mathbf{(q+G)\cdot r}}|\mathbf{k'}l' \rangle  \langle \mathbf{k'} l'|\mathrm e^{\mathrm{i}\mathbf{(q+G^\prime)\cdot r}}|\mathbf{k}l \rangle,\label{lfe_pol}
\end{eqnarray}
where $\mathbf{G}$ and $\mathbf{G^\prime}$ are arbitrary reciprocal lattice vectors. The dielectric function within LFE is given by the following matrix under RPA.
\begin{equation}
	\varepsilon_{\mathbf{G,G^\prime}}(\mathbf{q}, \omega) = \delta_{\mathbb{G,G^\prime}}- V(\mathbf{q+G})\Pi_{\mathbf{G}, \mathbf{G^\prime}}(\mathbf{q}, \omega),\label{lfe_eps}
\end{equation}
with the two-dimensional coulomb potential $V(\mathbf{q+G}) =2\pi e^2/\varepsilon_\mathrm{B}({\mathbf{q+G}}) $. The off-diagonal matrix elements in equation (\ref{lfe_eps}) give rise to LFE. If $G=G^\prime=0$, the dielectric matrix equation (\ref{lfe_eps}) reduces to the Lindhard dielectric function in equation (\ref{eps_dyn}). The optically detected macroscopic dielectric function is given by\cite{adler1962quantum,wiser1963dielectric}
\begin{equation}
\epsilon_M(\mathbf{q}, \omega) = \frac{1}{\varepsilon^{-1}_{G=0,G^\prime=0}(\mathbf{q}, \omega)},\label{mac_eps}
\end{equation}
where $\varepsilon^{-1}$ is the inverse of the matrix $\varepsilon_{\mathbf{G,G^\prime}}$. By comparing the macroscopic dielectric function $\epsilon_M(\mathbf{q}, \omega)$ to equation (\ref{eps_dyn}), one can know how LFE affect optical properties in a crystal\cite{louie1975local}. The energy loss function is formulated as
$S(\mathbf{q}, \omega) = -\mathrm{Im}(\frac{1}{\epsilon_M(\mathbf{q}, \omega)}) = -\mathrm{Im}[\varepsilon]^{-1}_{G=0,G^\prime=0}(\mathbf{q}, \omega) $.

\section{Optical properties and plasmons in non-twisted moir\'{e} structures} 

\subsection{Optical conductivity in graphene-based moir\'{e} structures}

\begin{figure}[h]
\includegraphics[width=0.9\textwidth]{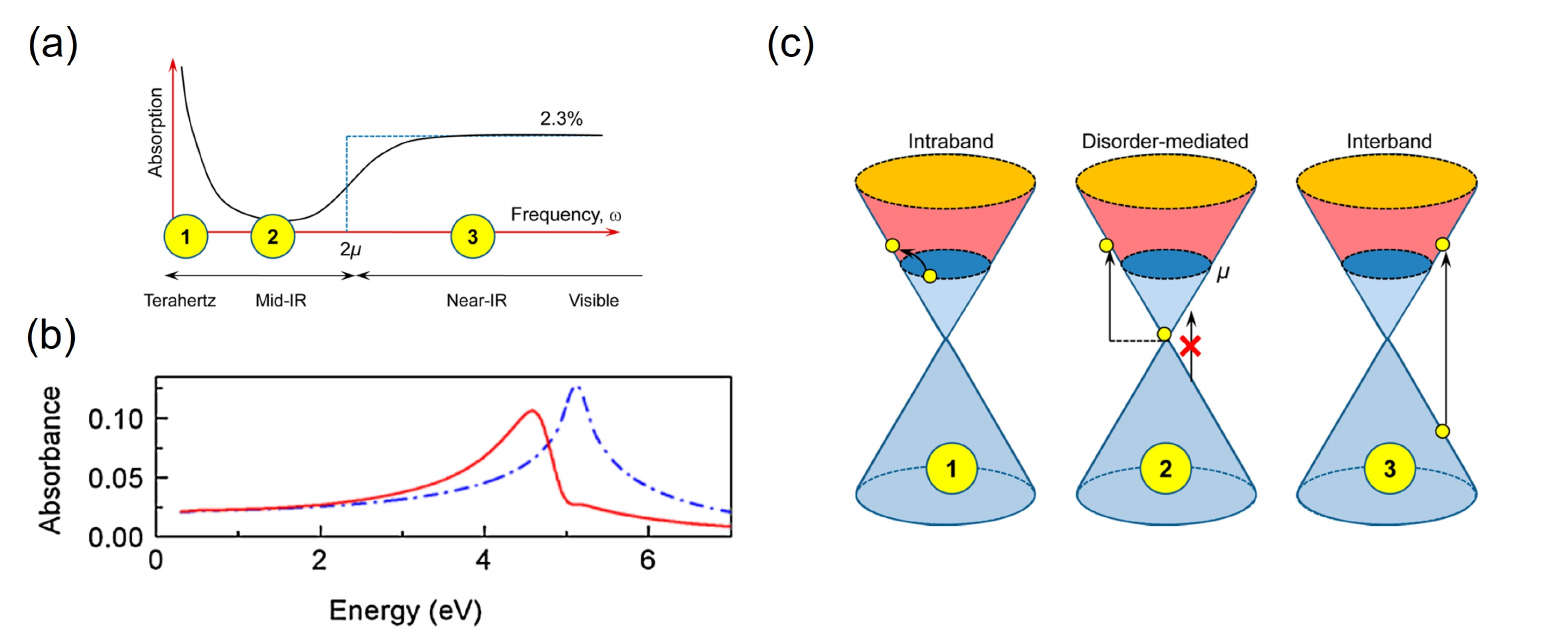}
\caption{Optical conductivity in monolayer graphene. (a) Illustration of a typical absorption spectrum of doped graphene. Reprinted with permission from \cite{Low2014graphene}. Copyright {2014} American Chemical Society. (b) First-principles absorbance of graphene with (red line)
and without (blue line) excitonic effects included. Reprinted with permission from \cite{yang2009excitonic} Copyright (2009) by the American Physical Society. (c) Illustration of the various optical transition processes.  Reprinted with permission from \cite{Low2014graphene}. Copyright {2014} American Chemical Society.}
\label{fig_op1}
\end{figure}

The optical properties of the graphene monolayer have many intriguing features\cite{koppens2011graphene,gonccalves2016introduction,xiao2016graphene}, such as a constant optical conductivity in the infrared regime and gate-dependent optical absorbance \cite{Low2014graphene}. As shown in Figure \ref{fig_op1}, firstly, there is a Drude peak at terahertz (THz) frequencies due to the intraband transitions. Secondly, for graphene with doping $\mu$, a minimal absorption in the mid-infrared frequencies occurs at finite $\omega < 2\mu$ due to Pauli blocking. Thirdly, a transition occurs around $\omega\approx 2\mu$ where direct interband processes lead to a constant optical conductivity $\sigma_0=\pi e^2/2h$. Finally, a sharp peak is located at 2$t$ (with $t$ the intralayer nearest neighbor hopping in graphene) arising from the interband transitions between the two van Hove singularities (VHS), which are logarithmically divergent points in electronic density of states (DOS) and corresponding to saddle points of band structure. This peak becomes red-shifted with a different line shape if we consider the electron-hole interaction \cite{yang2009excitonic}.


Graphene is usually supported on top of a hBN substrate to retain a high quality. 
When graphene is placed on the hBN substrate, a {\mo} pattern of 14 nm is formed in aligned samples due to the 1.8\% lattice mismatch between these two 2D materials \cite{xue2011scanning,wang2019new, wang2016gaps}. 
Undoubtedly, the periodic {\mo} potential induced in the graphene-hBN structure changes significantly the electronic structure  of graphene and leads to various novel quantum phenomena 
such as the emergence of the second-generation Dirac cones (located around some {\mo} energy $E_M$), the renormalization of the Fermi velocity, and a gap opening in the intrinsic Dirac cone \cite{wang2016gaps}. 

In the graphene-hBN {\mo} structure, there is a fast sublattice oscillation due to boron and nitrogen sites which results in a series of periodic potentials acting on the graphene monolayer~\cite{shi2014gate}. 
The first one is an scalar potential, which results from the {\mo} variation of the onsite terms, the second is a mass term originated from a local variation of the boron and nitrogen onsite terms, and a third one is a gauge potential resulting from the relaxation of the graphene atomic positions due to the presence of the hBN substrate~\cite{shi2014gate,Wallbank2013Generic}. The resulting potential, coupled to the electron pseudospin, can be probed directly through infrared spectroscopy, because optical transitions are very sensitive to wave functions of excited states. Consequently, in the experiment in Ref.~\cite{shi2014gate}, a remarkable absorption peak was detected around 2$E_M \sim 380$ meV
, which was only observed in the graphene-hBN heterostructure. Moreover, the absorption peaks were  found to be very sensitive to electron doping, 
which was revealed by a sharp decrease in its weight while increasing the electron concentration~\cite{shi2014gate}. The sharp drop could not be explained by the single-particle Pauli blocking effect whose energy was found to be small, 
but was due to a renormalization of the effective potential parameters induced by electron-electron interactions.  

\begin{figure}[t]
\includegraphics[width=\textwidth]{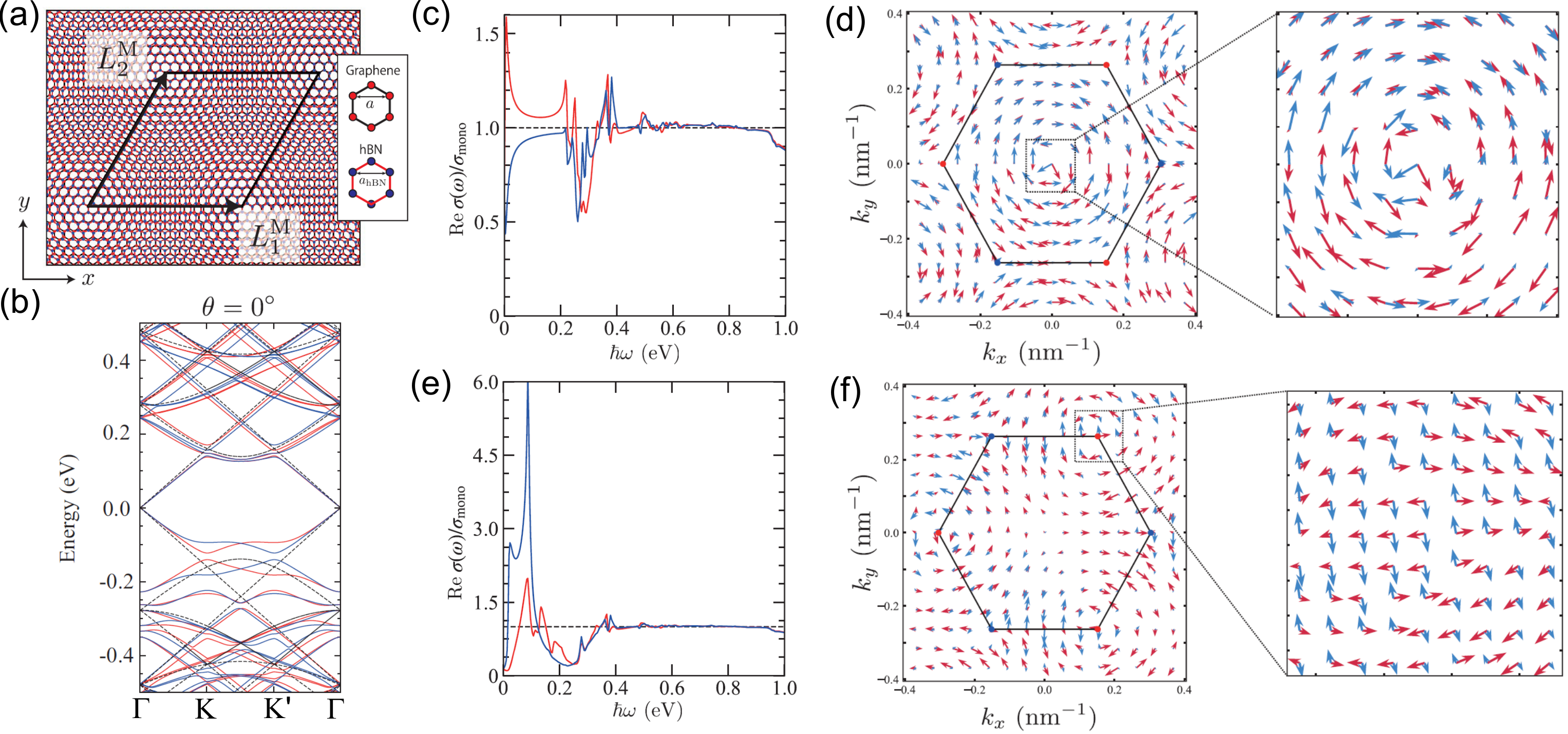}
\caption{Valley polarization of graphene-hBN structure under circularly polarized light irradiation. (a)  Crystal structure of graphene-hBN structure in real space for alignment case.  (b) The band structure of graphene-hBN in the continuum model. (c) Optical conductivity of graphene-hBN for K valley and Fermi level $E_F=0$ eV under circularly polarized light. The magenta (cyan) curve is the conductivity for LCP (RCP). (d) Distribution of dipole vectors for an interband transition near the
$E_F = 0.0$ eV in momentum space. The magenta (cyan) arrows are the real (imaginary) parts of the dipole vectors. The black hexagon is the
{\mo} BZ and the magenta (cyan) dots at the hexagon corners are the K (K') points. (e) The optical conductivity and (f) distribution of dipole vectors for $E_F = -0.13$ eV. Reprinted with permission from \cite{nakagahara2022enhanced} Copyright (2022) by the American Physical Society.}
\label{fig_op3}
\end{figure}

On the other hand, it is well-known that optical conductivity is typically dominated by the intraband Drude peak and interband transitions associated with singularities in the DOS. Theoretical works can provide a better understanding of which transitions among Bloch bands contribute to the optical conductivity. For instance, DaSilva~\cite{dasilva2015terahertz} employed a \textbf{k}$\cdot$\textbf{p} continuum Dirac model and the Kubo formula to investigate the optical conductivity of graphene aligned with hBN (as shown in Figure \ref{fig_op3}(a)). They discovered that the {\mo} pattern induced sharp THz peaks due to transitions between Bloch bands formed by the {\mo} potential. 
The particle-hole asymmetry of the {\mo} Bloch bands (see Figure \ref{fig_op3}(b)) was strongly reflected in the THz and IR conductivity, which was always Drude-dominated when the Fermi level lied above the Dirac point, but it was interband-dominated when the Fermi energy lies in a relatively narrow interval below the Dirac point. In addition, Abergel~\cite{abergel2015infrared} suggested that a study of the absorption spectra as a function of the doping for an almost completely full first miniband was necessary to extract meaningful information about the {\mo} characteristics from optical absorption measurements as well as to distinguish between various theoretical proposals for the physically realistic interactions between graphene and hBN. One of the main findings on Ref.~\cite{abergel2015infrared} was the fact that distinct {\mo} perturbations can result in similar absorption spectra. 

The effect of polarized light was studied in Ref.~\cite{nakagahara2022enhanced}. It was found that a broken spatial symmetry in the graphene-hBN structure may induce valley polarization, which could be investigated by measuring the optical conductivity under circularly polarized light irradiation, as shown in Figures \ref{fig_op3}(c) and (e). The conductivity depended on the direction of rotation of the circularly polarized light, especially in the infrared and terahertz regions. In particular, for a photon energy less smaller than 0.1 eV, the difference between left-handed circularly polarized light (LCP) and right-handed circularly polarized light (RCP) became larger. In this energy region, the interband transition from valence to conduction bands dominated. The real and imaginary parts of the dipole vectors are orthogonal at the $\Gamma$ point. Thus, the valley-selective circular dichroism (valley polarization) was induced by the irradiation of circularly polarized light, and was responsible for the states near the $\Gamma$ point (Figure \ref{fig_op3}(d)). For hole doping case, in the region of $\omega < 0.1$ eV, the difference in the optical conductivities between LCP and RCP became larger. The real and imaginary parts of the dipole vectors are mutually orthogonal at the K and K' points, which were responsible for the valley-selective circular dichroism. In fact, the {\mo} potential of aligned graphene-hBN structures can be tuned by a twist angle that continuously change optical intraband and interband transitions in graphene-hBN {\mo} structures\cite{liu2023tunable}.

\subsection{Plasmons in  graphene-based moir\'{e} 
structures}

The plasmons in graphene-based heterostructures \cite{dean2012graphene} have attracted a lot of attentions due to the fact that plasmons in pristine graphene have very promising perspectives\cite{Hwang2007dielectric,Eberlein2008plasmon}. 
In this section, we will mainly review plasmonic properties in graphene-hBN and graphene-metal {\mo} structures. 

\begin{figure}[h]
\includegraphics[width=1\textwidth]{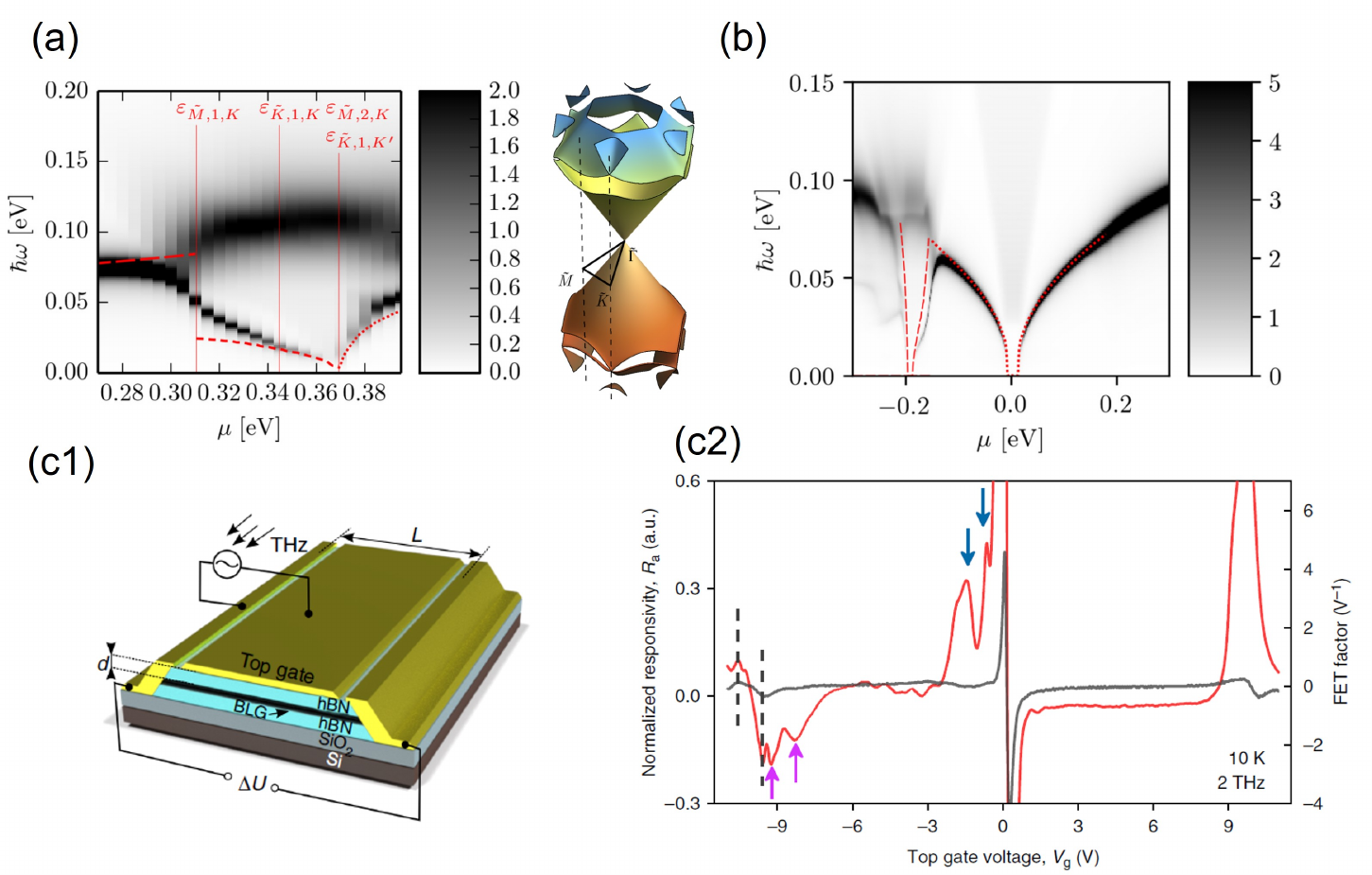}
\caption{Plasmon in graphene-hBN moir\'{e} structures. (a) Theoretical predicted plasmon energy relation versus hole-side chemical potential in a density plot of loss function within LFE (left-hand side) and corresponding miniband structure of graphene-hBN (right-hand side). The vertical thin solid lines in the density plot denote the doping levels crossing diferent band edges in the miniband structure.Reprinted with permission from \cite{Tomadin2014generation} Copyright (2014) by the American Physical Society. (b) Particle-hole asymmetric plasmon emerging in a density plot of loss fucntion within LFE for graphen-hBN moir\'{e} structure. Reprinted with permission from \cite{Tomadin2019plasmons} Copyright (2019) by the American Physical Society. (c1) Illustration of the encapsulated bilayer graphene-hBN field effect transistor. (c2) Plasmons from miniband transition in bilayer graphene-hBN structure. Reprinted from \cite{Bandurin2018resonant} CC BY 4.0.} \label{grasupplas}
\end{figure}

As mentioned in section 3.1, stacking graphene on a hBN substrate allows to engineer the electronic band structure of graphene by the induced moir\'{e} potential.
An important consequence is the emergence of satellite Dirac points in the moir\'{e} minibands~\cite{yankowitz2012emergence} since they could give rise to collective excitations that are different from pristine graphene.
In 2014, a theoretical study on plasmons in graphene-hBN unveiled that new plasmon modes can be generated due to transitions between satellite Dirac points, as shown in Figure \ref{grasupplas}(a) \cite{Tomadin2014generation}. 
The K-point and M-point plasmon modes (dotted and short-dashed lines in Figure \ref{grasupplas}(a), respectively), appeared alongside a Dirac plasmon mode (long-dashed line) in hole-doped graphene. 
Based on a continuum model and including the LFE, further calculations demonstrated a dramatic asymmetry of the plasmon dispersion at positive and negative potentials, as seen in Figure \ref{grasupplas}(b), and also predicted several plasmon modes arising from interband transitions between minibands. 
Experimentally, the measured optical response using a s-SNOM tip in moir\'{e}-patterned graphene was enhanced with respect to pristine graphene\cite{Ni2015plasmons}. A composite plasmon in graphene-hBN moir\'{e} structures was also observed, originating from intraband transitions near the Fermi energy and predicted interband transitions corresponding to structure mini-bands \cite{Ni2015plasmons}. Nevertheless, up to date, the calculated terahertz plasmon from \cite{Tomadin2014generation} and asymmetry plasmon from\cite{Tomadin2019plasmons} have not been observed experimentally. In addition, based on an antenna-mediated coupling of a bilayer graphene (BLG) field-effect transistor device, shown in Figure \ref{grasupplas}(c1), miniband plasmons in BLG-hBN were also observed (see Figure \ref{grasupplas}(c2)).


\begin{figure}[thpb]
\includegraphics[width=1\textwidth]{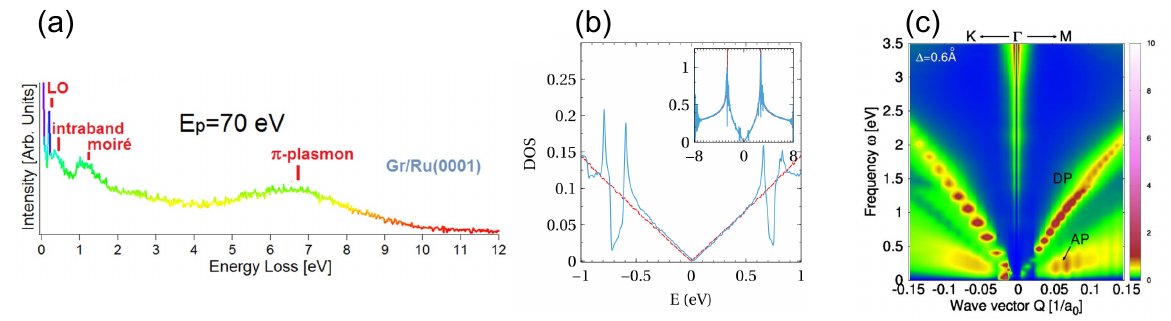}
\caption{Plasmon in graphene-metal moir\'{e} structures. (a) High-resolution electron energy loss specstroscopy of graphene on Ru(0001), aquired with an energy of the primary electron beam $E_p$ of 7 eV. The longitudinal optical (LO) phonon, intraband plasmon, {\mo} plasmon and $\pi$-plasmon are observed. (b) moir\'{e} VHS in graphene-Cu moir\'{e} structures from density functional theory calculations. Used with permission of IOP Publishing, Ltd, from \cite{Politano2017effect}; permission conveyed through Copyright Clearance Center, Inc. (c) The loss function intensity plot showing an acoustic plasmon (AP) predicted in graphene-metal structure doped with alkali-metal atoms. Reproduced with permission from \cite{Despoja2019strong} Copyright (2019) by the American Physical Society.}
\label{grametalplas}
\end{figure}

The observation of miniband plasmons in graphene-hBN also motivated the exploration of electronic excitations in graphene-metal structures. The moir\'{e} potential was induced when Cu atoms were deposited on graphene, forming a moir\'e structure, giving rise to extra VHS associated with minibands (see Figure \ref{grametalplas}(b)). This resulted in a moir\'{e} plasmon mode with energy $\sim$ 1.5 eV, which was contributed by interband transition between VHS (as shown in Figure \ref{grametalplas}(a)). The existence of this plasmon mode was theoretically confirmed by using the TBPM method \cite{Politano2017effect}. Interestingly, when the moir\'{e} potential induced in chemically doped graphene on the Ir(111) metallic surface was suppressed, plasmon could be still significantly modified, generating an acoustic plasmon (AP) mode along with an intraband Dirac plasmon (DP) mode (see figure \ref{grametalplas}(c)) \cite{Despoja2019strong}. This AP was induced by the screening effect of metallic materials or graphene-metal hybridization rather than by the moir\'{e} reconstruction and had also been widely studied in previous works\cite{politano2011evidence,politano2012effects,langer2011sheet}. 

\section{Optical properties and plasmons in twisted {\mo} structures}

\subsection{Optical conductivity of twisted bilayer graphene} 

The recent discovery of correlated electronic states and superconductivity in TBG~\cite{andrei2021marvels,park2021tunable,cao2018unconventional} has sparked a great interest in twisted moir\'{e} systems. In TBG the interlayer interactions induce significant distortions in the low-energy bands. This leads to distinctive electronic effects that differ from those observed in non-twisted graphene systems. At low angle, the interference between the moir\'{e} periods produces a long wavelength moir\'{e} pattern~\cite{dos2007graphene,shallcross2008quantum, shallcross2010electronic,Bistritzer2011moire,trambly2010localization,mele2010commensuration, dos2012continuum}. Characteristic properties like VHS and band gaps become evident in the infrared region~\cite{moon2013optical} and theoretical works~\cite{dos2007graphene,shallcross2008quantum, shallcross2010electronic,Bistritzer2011moire,trambly2010localization,mele2010commensuration, Trambly2012numerical,dos2012continuum} have demonstrated that the moir\'{e} patterns in TBG can give rise to narrow bands that largely contribute to the correlated effects observed in this system~\cite{park2021tunable,cao2018unconventional,yankowitz2019tuning,lu2019superconductors,polshyn2019large,sharpe2019emergent,serlin2020intrinsic,chen2020tunable,saito2020independent,zondiner2020cascade,wong2020cascade,stepanov2020untying,xu2020correlated,Choi2021correlation,rozen2021entropic,cao2021nematicity,stepanov2021competing,Oh2021evidence,Xie2021fractional,berdyugin2022out,turkel2022orderly}. 

\begin{figure}[thpb]
\includegraphics[width=1\textwidth]{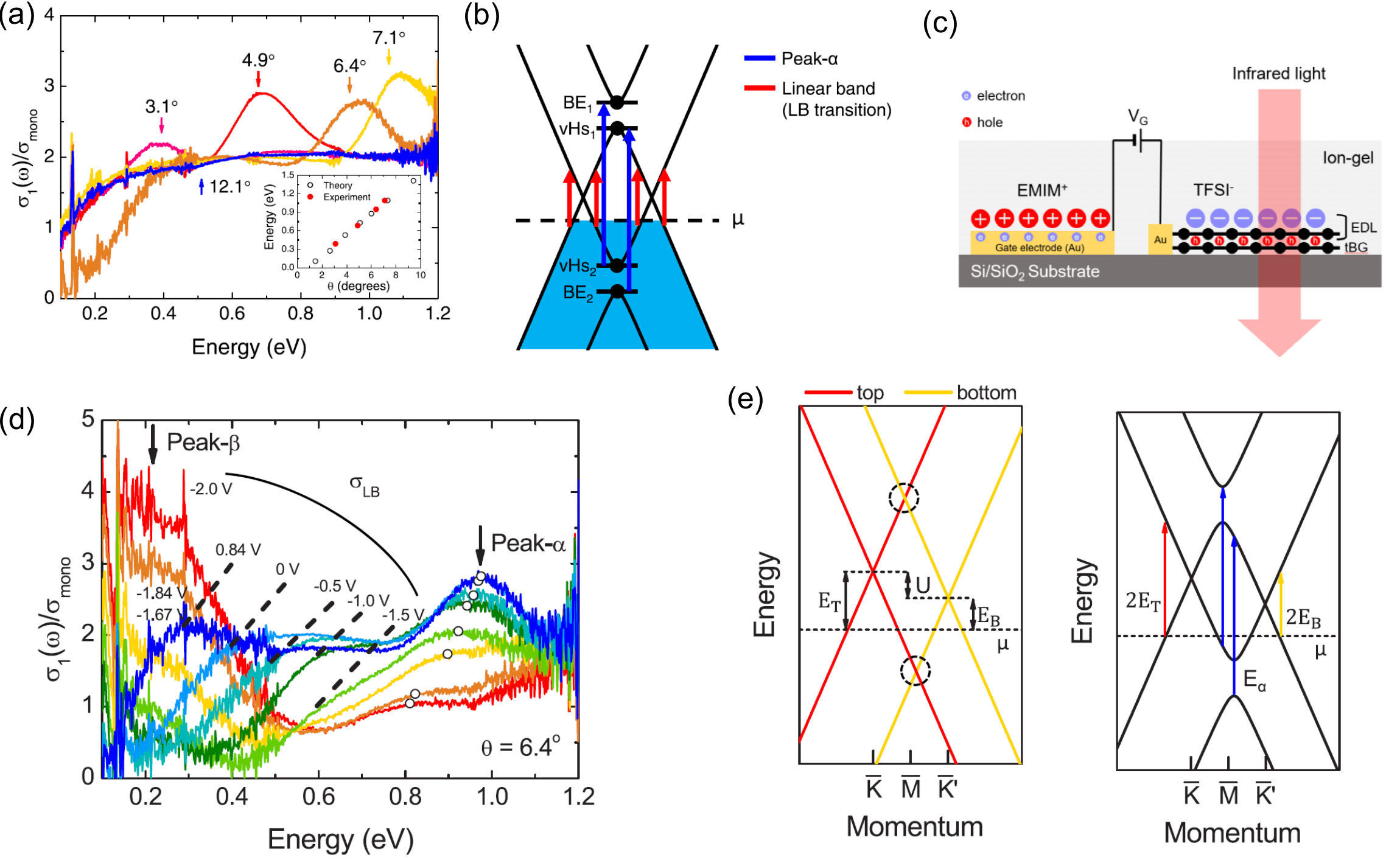}
\caption{Gate tunable optical absorption of TBG. (a) Optical conductivity $\sigma_1$ of five TBG samples with five different twist angles. $\sigma_{\mathrm{mono}}$ is the optical conductivity of monolayer graphene. (b) Electronic band diagram of TBG. BE and vHs stand for the band edge of the second band and the saddle-point van Hove singularity, respectively. There exist two kinds of optical transitions as indicated by the red and blue arrows. (c) Schematic view of the ion-gel gating circuit and the infrared transmission measurement. EMIM and TFSI are ionic liquids. (d) Optical conductivity of TBG with various gate voltages $V_G$. The twist angle is $\theta=6.4^\circ$. (e) (Left side) The band structure of TBG under gating. The top band and bottom band shift by $E_T$ and $E_B$, respectively. $U=E_T - E_B$ is their difference. Here, the gap opening is omitted for clarity. (Right side) Optical transitions of the gated TBG. Reprinted with permission from \cite{Yu2019gate} Copyright (2019) by the American Physical Society.}
\label{fig_op4}
\end{figure}

An interesting feature of TBG is that the VHS can be moved to arbitrarily low energies by modifying the twist angle. One interesting optical analysis of TBG was performed by Yu \emph{et. al}.~\cite{Yu2019gate}. In the experiment, different optical conductivities were obtained by varying the twist angle. As shown in Figure~\ref{fig_op4}(b), the low-energy optical spectrum of TBG was characterized by a linear-band (LB) absorption (pointed out by the red arrows). This was an indication that the interlayer interaction hybridized the LBs of the two monolayers and, as a consequence, there were two isolated bands with an avoided crossing with the remote bands. In Figure~\ref{fig_op4}(b), the transition between the saddle-point VHS$_2$ $\to$VHS$_1$ was forbidden by the lattice symmetry \cite{moon2013optical}. However, the transition between VHS and the band edge (BE) of the second band exhibited prominent peaks (peak-$\alpha$ indicated by the blue arrows). Two interesting features, shown in Figure \ref{fig_op4}(a), were found in this experiment: firstly, a frequency-independent conductivity, $2\sigma_0$, which came from the LB transition, and secondly, an angle-dependent peak-$\alpha$ resulting from the transitions between VHSs and BE. Interestingly, the peak-$\alpha$ was blueshifted as the twist angle $\theta$ increased, revealing a dependence on the twist angle. Furthermore, by designing an ion-gel gating circuit (in Figure \ref{fig_op4}(c)) it was possible to investigate the gating effect on the properties of these devices.  It was found that in the absorption profile, cf. Figure~\ref{fig_op4}(d), (i) the absorption edge of $\sigma_{LB}$ had a broadening and was shifted to higher energy, and (ii) the peak-$\alpha$ was shifted to a lower energy with a reduced intensity. In addition, a modification of the band structure with gating was found, because in the presence of a vertical electric field, Figure \ref{fig_op4}(e), the Dirac cones of each monolayer shifted in opposite directions. This was theoretically described in~\cite{SanJose2013Helical} and recently in~\cite{tsim2020Perfect}, allowing to modify the optical transitions~\cite{Moon2014optic}.

Additional evidence of the existence of isolated narrow bands with an enhanced density of states was reported in 2013 by Zou et al.~\cite{zou2013terahertz}. By means of a terahertz time-domain spectroscopy, the optical conductivity of TBG was obtained at different temperatures in the frequency range 0.3–3 THz. One of the main findings in this work was a Drude-like response with a strong peak in $\sigma_1(\omega)$ at $\sim$ 2.7 THz, which was identified as peak-$\alpha$ in TBG with $\theta=1.16^\circ$. 

\begin{figure}[thpb]
\includegraphics[width=0.95\textwidth]{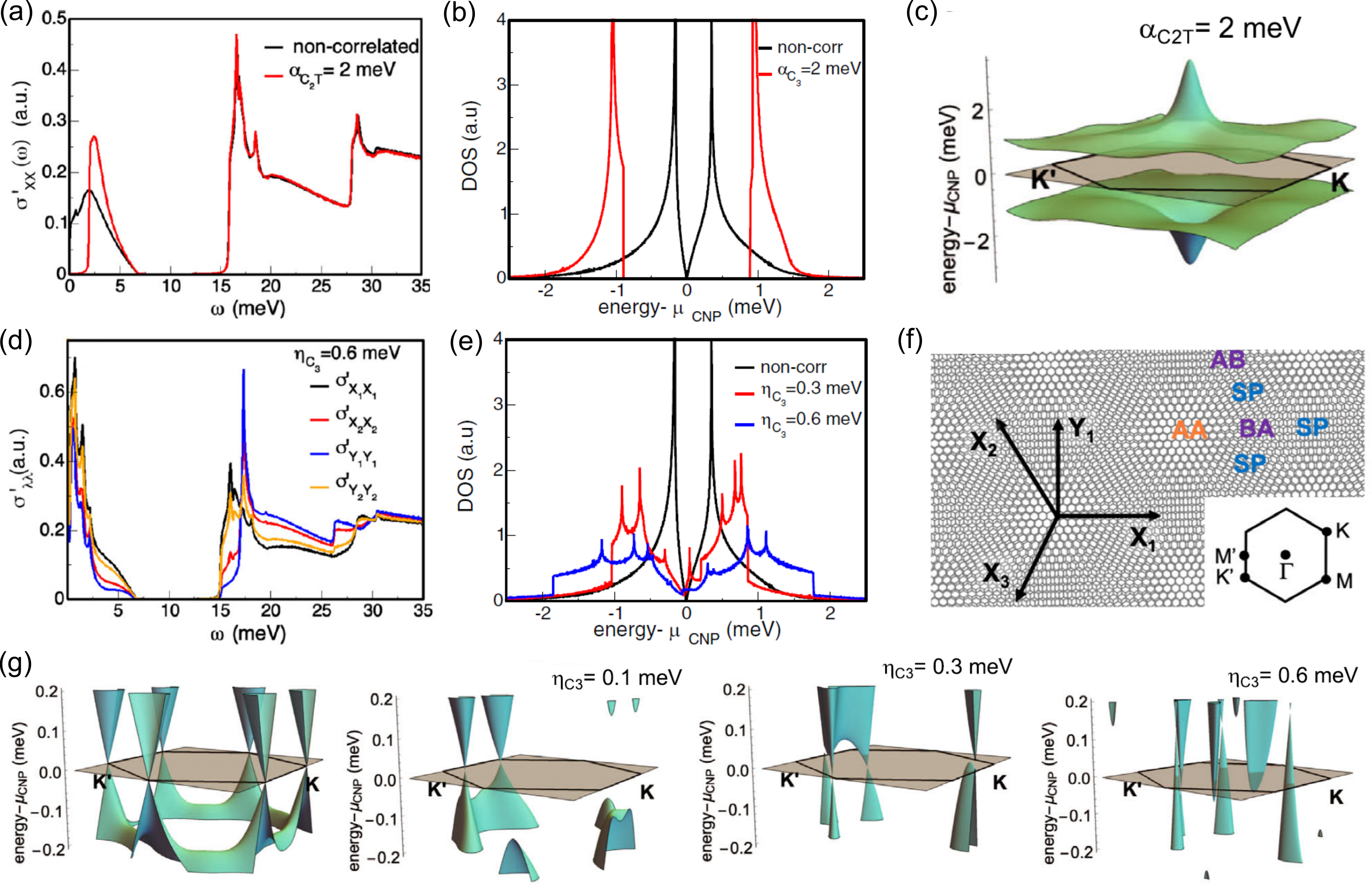}
\caption{Optical conductivity, DOS and band structures of undoped TBG with different symmetry breaking orders. (a) Optical conductivity, and (b) low energy DOS of the TBG at the CNP in a $C_2T$ symmetry breaking state $\alpha$ and in the non-correlated state. (c) Flat bands of the TBG in the $C_2T$ symmetry breaking state. (d) Optical conductivity and (e) DOS for the nematic state $\eta$. The optical conductivity in (d) is plotted along four different directions, illustrated in (f). (f) Sketch of the $C_3$ related $X_i$ directions and of $Y_1$, and the BZ of TBG with the corresponding symmetry points. (g) Zoom of the flat band structure for the TBG with different values of $\eta_{C_3}$. Note that the optical conductivity in (a) along the four directions are equal. Reprinted from \cite{Calderon2020correlated} CC BY 4.0.}
\label{fig_op5}
\end{figure}

Interestingly, TBG shows an angle-dependent optical conductivity, which could be utilized to characterize the twist angle. For example, Sunku et al.~\cite{sunku2020nano} combined nano-photocurrent and infrared nanoscopy methods, which enabled access to the local electronic phenomena at length scales as short as 20 nm, and identified domains of varying local twist angles. In addition, Calder\'on~\cite{Calderon2020correlated} reported that the optical conductivity measurements could be used to distinguish different symmetry breaking states, and may reveal the nature of the correlated states in the flat bands that appear in TBG. As shown in Figures \ref{fig_op5}(a)-(c),   
in a correlated order which breaks the $\mathcal{C}_2\mathcal{T}$ symmetry, named $\alpha$ here, a gap was opened at the Dirac points in K, resulting in a reorganization of the spectral weight. In TBG without correlations or external symmetry breaking, the lattice had $\mathcal{C}_3$ symmetry, Figure~\ref{fig_op5}(f). Moreover, in a system with $\mathcal{C}_3$ symmetry$\sigma_{X_iX _j}=\sigma_{Y_iY_j}$. Here, $X_i$ is the direction, as illustrated in Figure \ref{fig_op5}(f). Therefore, the optical conductivity along the four directions ($X_1X_1$, $X_2X_2$, $Y_1Y_1$ and $Y_2Y_2$) were equal in the correlated order, but different in the reported nematic state (named $\eta_{C_3}$, which lowers the rotational symmetry of TBG). Furthermore, in the nematic state, when the flat band was partially filled, the DOS was modified. With larger values of the amplitude of the order parameter $\eta_{C_3}$, the Dirac points moved away from the charge neutrality point and hole and electron Fermi pockets were generated. Additional Fermi pockets appeared, leading to new band crossings between the lower and upper flat bands (Figure \ref{fig_op5}(g)).   

On the theoretical side, we highlight the following works: in Ref. \cite{tabert2013optical}, the authors used a continuum model to study the frequency-dependent conductivity at different levels of chemical potential. Moon et al.~\cite{moon2013optical} performed both tight-binding and continuum calculations of the optical conductivity, and analytically explained the optical selection rules in terms of the symmetry of the effective Hamiltonian. In addition, Stauber et al.~\cite{Stauber2013Optical} calculated chemical potential dependent Drude weight of the optical conductivity in TBG by means of a continuum model. The excitonic effects, for instance, electron-hole interactions \cite{Havener2014vanhove}, and the self-consistent Hartree potential \cite{Novelli2020optical}, were also investigated in the optical spectra of TBG. It was found that in TBG under strain~\cite{Dai2021effects}, the peaks associated with transitions between the flat bands in the optical conductivity were highly sensitive to the direction of the strain. The effects of a magnetic field~\cite{moon2013opticalmagne} and magnetic impurities on the optical conductivity~\cite{natalin2023optical}, as well as optical activity in TBG have also been analysed\cite{Chang2022theory}. Analysis in stacking configurations~\cite{wang2010stacking,vela2018electronic}, quantum dots and large twist angles ($\theta\geq5^\circ$)~\cite{Wang2021polarization} reveals additional optical properties.

\subsection{Optical properties of other twisted 2D materials}

\begin{figure}[thpb]
	\includegraphics[width=1\textwidth]{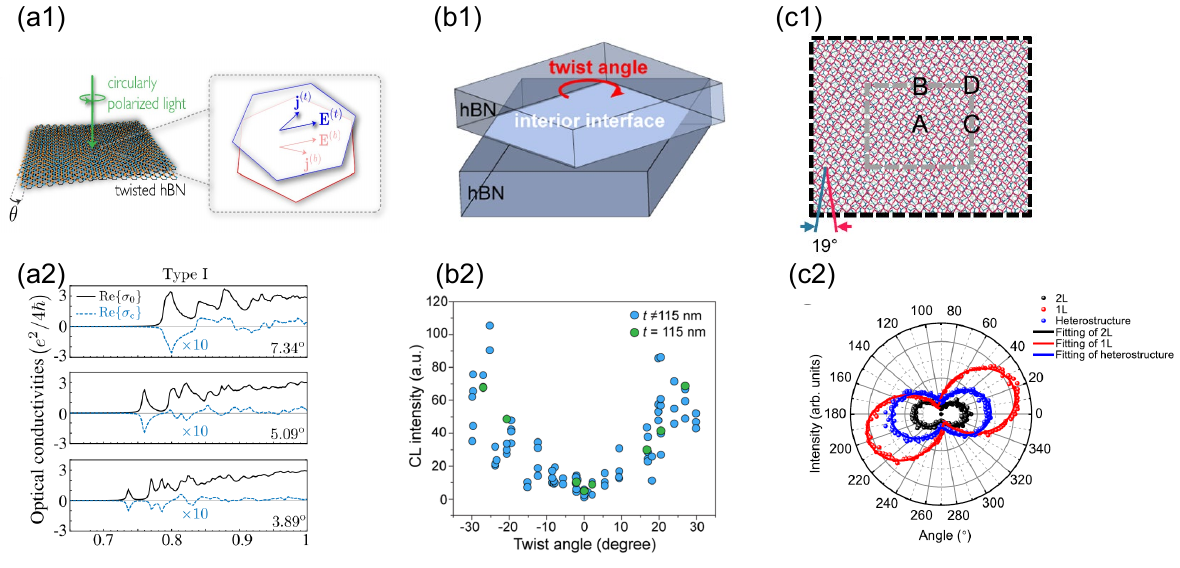}
	\caption{Optical properties of other twisted 2D materials. (a1) Illustration of twisted bilayer hBN with incident circularly light. (a2) Real part of total (black lines) and chiral (blue lines) optical conductivity. Reprinted with permission from \cite{Ochoa2020flat} Copyright (2020) by the American Physical Society.
 (b2) Twist-angle dependent cathodoluminescence (CL) intensity in (b1) twisted hBN multilayers. Reprinted with permission from \cite{Lee2021tunable} Copyright {2021} American Chemical Society. (c2) Anisotropic photoluminescence emissions in (c1) twisted monolayer/bilayer phosphorene heterostructure with twist angle $19^{\circ}$ comparing to monolayer (1L) and bilayer phosphorence (2L). Reprinted with permission from \cite{Zhao2021anisotropic}.}\label{twhbnopt}
\end{figure}

Owing to the chiral symmetry, twisted bilayer hBN displays circular dichroism, which has a different absorption of left and right circularly polarized light~\cite{Wu2019Chiral}. This property can be tuned by stacking and twisting~\cite{Ochoa2020flat}. The circular dichroism is proportional to the ratio of chiral conductivity to the total conductivity $\sigma_0$ (shown in Figure \ref{twhbnopt}(a2)). The chiral response indicated that twisted bilayer hBN had different absorption to left and right-polarized light. Besides twisted bilayer hBN, twisted hBN films have also attracted recent experimental research, in particular Lee \emph{et al.}\cite{Lee2021tunable} found that both wavelength and intensity of luminescence were tunable. These properties were found to be enhanced with the twist angle between the hBN interface layers increased, as seen in Figure \ref{twhbnopt}(b2). The origin of this enhancement was that the moir\'{e} sub-band gap decreased with twist angle.  This experiment indicates that the moir\'e potential is relevant in the moir\'{e} structures composed of bulk-like materials . 

On the other hand,  moir\'{e} optical properties in twisted semiconductors have also been investigated. For example, in the anisotropic twisted monolayer/bilayer phosphorene heterostructure, shown in Figure \ref{twhbnopt}(c1). The detected anisotropic optical transitions were notably different from the optical features of the corresponding monolayer and multilayer phosphorene\cite{qiao2014high,li2017direct}, even at a large angle like 19$^\circ$, as illustrated in Figure \ref{twhbnopt}(c2). The reason behind this effect is that the moir\'{e} potential resulted in a strong hybridization between the twisted layers. Furthermore, the optical moir\'{e} transitions were sensitive to the twist angle\cite{Zhao2021anisotropic}. Additionally, the twisted heterostructure of anisotropic materials such as black phosphorus and orthorhombic molybdenum trioxide can be used to control light polarization state~\cite{Khaliji2022twisted}. Twisting large angles such as $\theta =21.81^\circ$ and $\theta =32.22^\circ$ can also serve as a way to reduce interlayer interaction of bilayer MoS$_2$, which can induce a higher value of absortance than untwisted case~\cite{lee2023achieving}.


\begin{figure}[thpb]
	\includegraphics[width=1\textwidth]{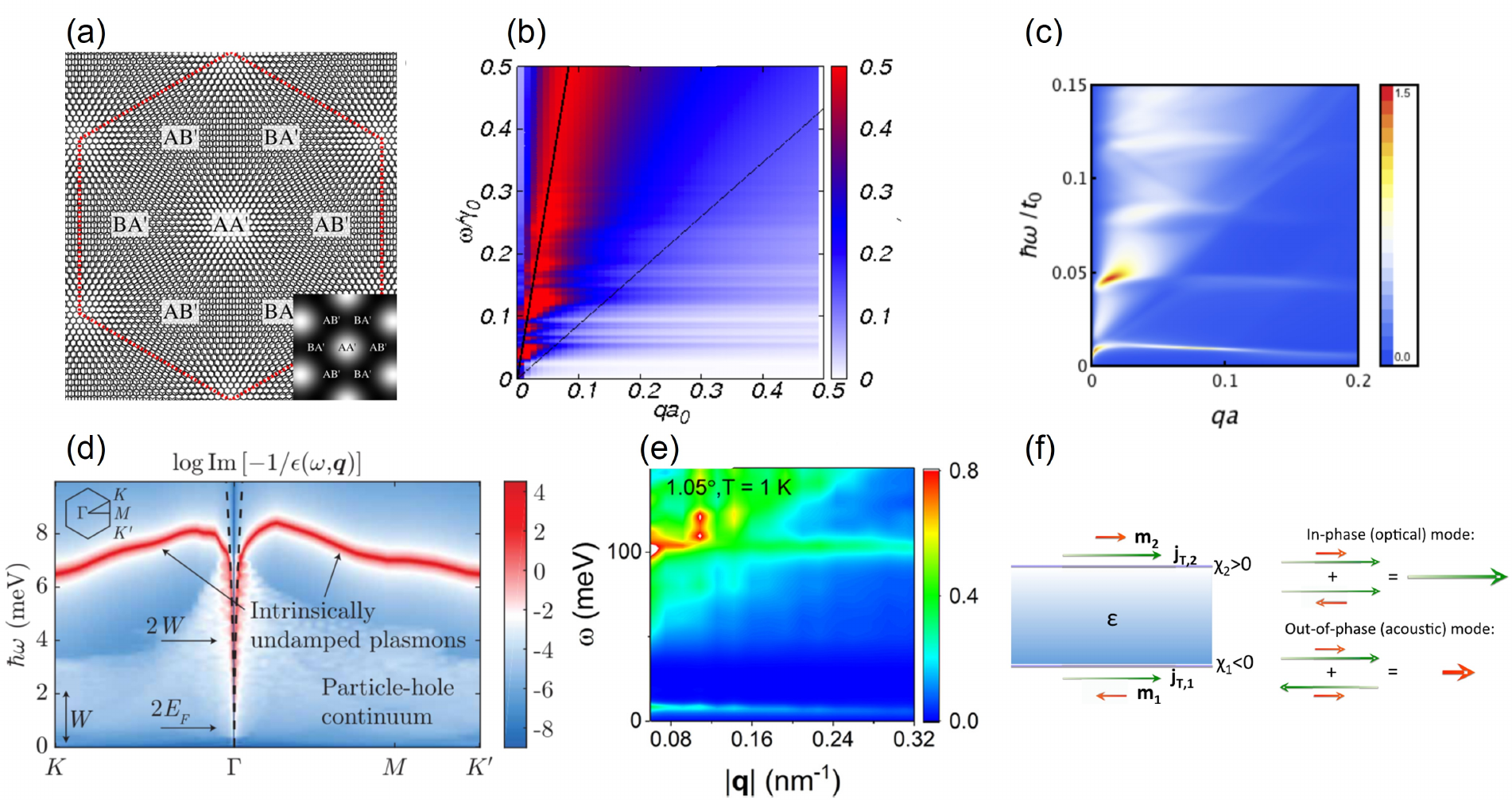}
	\caption{Theoretical exploration of plasmons in twisted bilayer graphene. (a) Moir\'{e} structure of twisted bilayer graphene. The moir\'e pattern contains AA, AB and BA high-symmetry stackings. (b) Linear plasmon emerging in the loss function intensity plot of magic-angle twisted bilayer graphene. Reprinted from \cite{Stauber2013Optical}. (c) Density plot of the loss function showing quasi-flat plasmons in twisted bilayer graphene at $1.61^\circ$ Reprinted with permission from \cite{stauber2016plas} Copyright {2016} American Chemical Society. (d) Undamped plasmons in doped magic-angle twisted bilayer graphene. $W$ is the width of the narrow bands. $E_F$ is the Fermi energy. Reprinted from \cite{Lewandowski2019intrinsically}. (e) Dispersion-less and low-damped plasmon appearing in the loss function spectrum of undoped magic-angle TBG  at temperature T = 1K. Reprinted with permission from \cite{Kuang2021collective} Copyright (2021) by the American Physical Society. (f) Chiral plasmon response in twisted bilayer graphene. Reprinted with permission from \cite{Stauber2018chiral} Copyright (2018) by the American Physical Society.}
	\label{caltbgplas}
\end{figure}

\subsection{Plasmons of twisted bilayer graphene}
Twisted bilayer graphene offers new degrees of freedom on tuning the electromagnetic response, for example, the twist angle~\cite{Yin2016Selectively, Havener2014vanhove, Patel2015Tunable}. Plasmons are collective charge oscillations that lead to nanoscale optical fields. One of the pioneer works was that of Stauber \emph{et al.}~\cite{Stauber2013Optical} who theoretically investigated the plasmonic spectrum of TBG via a continuum model. They found that the TBG interlayer coupling gave rise to a finite Drude weight, even in the undoped case. This allowed for the existence of plasmons that was weakly Landau-damped due to the quasi-localized nature of the interband transition states. As shown in Figure \ref{caltbgplas}(b), acoustic interband plasmon modes appeared at zero chemical potential and changed to conventional $\sqrt{q}$ modes with non-zero doping levels in the first magic-angle ($1.05^{\circ}$) TBG~\cite{Stauber2013Optical}. Moreover, plasmon modes in TBG were dependent on both twist angle and the chemical potential. Interestingly, quasi-flat plasmon modes and renormalized Fermi velocity (approached zero) were predicted in TBG even for twist angles ($\sim 1.61^\circ$) larger than the magic angle, as seen in Figure \ref{caltbgplas}(c). These collective excitations were explained as oscillation of localized states around the AA regions (see Figure \ref{caltbgplas}(a))~\cite{stauber2016plas}. Moreover, intrinsically undamped and quasi-flat plasmon modes were discovered in doped magic-angle graphene, as depicted in Figure \ref{caltbgplas}(d)~\cite{Lewandowski2019intrinsically}. Conversely, at zero doping, including the effects of atomic relaxation, low-damped and damped plasmons were observed to emerge in the magic-angle configuration\cite{Kuang2021collective}. A further theoretical study found that in the long wavelength limit, the plasmon energy could be independent of doping level, but can be tuned by the bias voltage in magic-angle TBG\cite{Ding2021corrugation}. 
These findings distinguish the TBG system from the 2D electron gas (2DEG) that has a traditional $\sqrt{q}$ plasmon dispersion with energy dependent on charge density\cite{Stern1967polarizability}. Moreover, other unusual plasmon features such as plasmonic Dirac cone and plasmon non-reciprocity, were discovered in biased magic-angle TBG~\cite{Brey2020plasmonics,Zhou2021carrier,Papaj2020plasmonic,Song2021electric}. In addition, a further design of a TBG device as in Figure \ref{caltbgplas}(f), allowed to excite chiral longitudinal plasmonic modes with different phases. Additional chiral responses and plasmon edge states were exploited and were found to enhance the electromagnetic near-fields chirality in TBG~\cite{Lin2020chiral,Stauber2020plasmon,Margetis2021theroy}. Furthermore, theoretical studies explored how plasmon in TBG were influenced by electron-electron interaction\cite{Novelli2020optical,Ding2022role}, finite size of TBG\cite{westerhout2021quantum}, and magnetic field\cite{Do2023magnetop}. 

\begin{figure}[thpb]
	\includegraphics[width=1\textwidth]{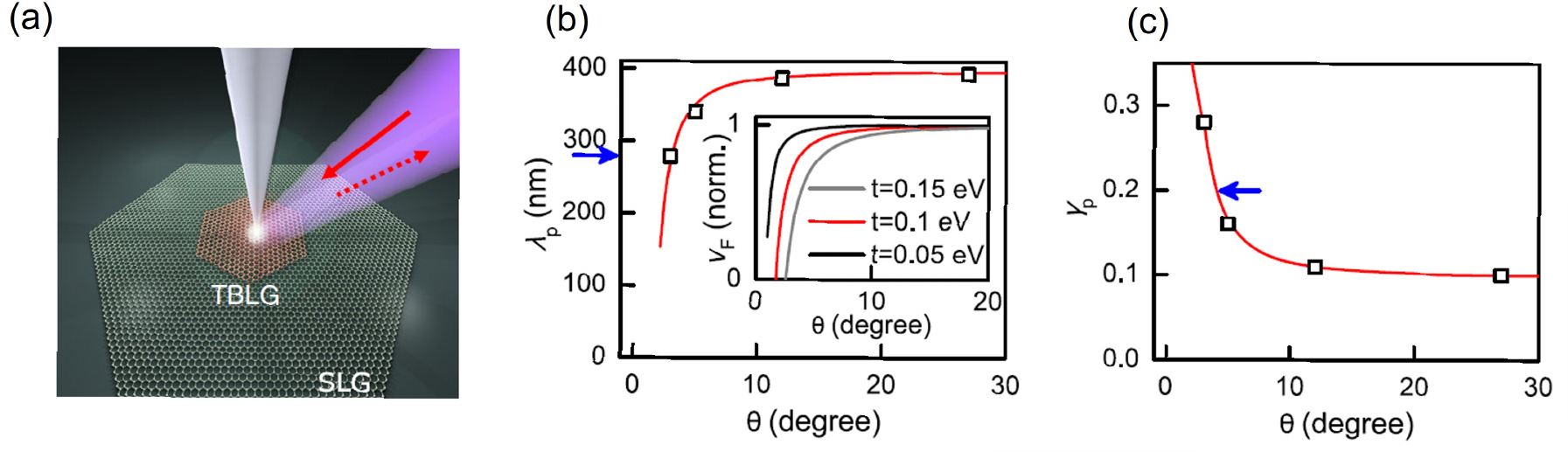}
	\caption{Experimental investigation of plasmons in twisted bilayer graphene. (a) Illustration of the nanoinfrared imaging experiment using s-SNOM in TBG using laser energy of E=0.11 eV. (b) Plasmon wavelength of TBG versus twist angles. The insets plots the normalized Fermi velocity of TBG at different interlayer coupling energy t. (c) plasmon damping rate at different twist angles. The blue arrows in (b) and (c) mark the value of plasmon wavelength and damping rate in single layer graphene, respectively. Reprinted with permission from \cite{Hu2017realspace}. Copyright (2017) by the American Physical Society.}
	\label{expltbgplas}
\end{figure}

Experimentally, the plasmon wavelength and damping rate were investigated by infrared s-SNOM with an excitation energy of 0.11 eV~\cite{Hu2017realspace}. It was found that TBG with decreased twist angles leaded to the decrease of the plasmon wavelength, shown in Figure \ref{expltbgplas}(b), reflecting a renormalization of the Fermi velocity of the Dirac fermions at different twist angles. 
A reduced Fermi velocity is attributed to the enhanced interlayer interaction at twist angles, giving rise to a relaxation of the plasmon wavelength. However, the plasmon damping rate was found to be smaller with larger twist angles, as seen in Figure \ref{expltbgplas}(c), likely due to stronger charge scattering rates \cite{Hu2017realspace}. On the other hand, the propagation of plasmon polaritons was studied by infrared nano-imaging in TBG~\cite{Sunku2018photonic}. A linear-like plasmon mode (that is, the dispersion of a plasmon mode is closely linear) of $1.35^{\circ}$ TBG was observed by Hesp \emph{et al.} \cite{Hesp2021observation}. The plasmon mode had an energy around 220 meV, due to interband collective excitations whose spectrum was reproduced with a continuum model with reduced AA tunnel coupling.   

In addition, chiral plasmons, the surface electromagnetic waves showing non-reciprocal propagation, in TBG were experimentally reported in Ref.~\cite{Huang2022observation}. They are achieved due to the uncompensated Berry flux of the electron gas under optical pumping. They were found to be characterized by two peaks appearing in the extinction spectra. These low-energy plasmon modes arose from interband transition with broken time-reversal and inversion symmetry. In the experiment, a plasmonic mode whose group velocity approaching to zero (termed slow plasmon modes) was identified around 0.4 eV, which stemmed from interband transition between subbands in lattice-relaxed AB domains \cite{Huang2022observation}, compared to theoretically predicted quasi-flat plasmon mode generated in the AA regions\cite{stauber2016plas}. These slow plasmon modes could couple to light and form slow surface plasmon polaritons, which also provide potential for constructing optical metamaterials\cite{Su2022atomic}.

\subsection{Plasmons of twisted multilayer graphene}
\begin{figure}[thpb]
	\includegraphics[width=1\textwidth]{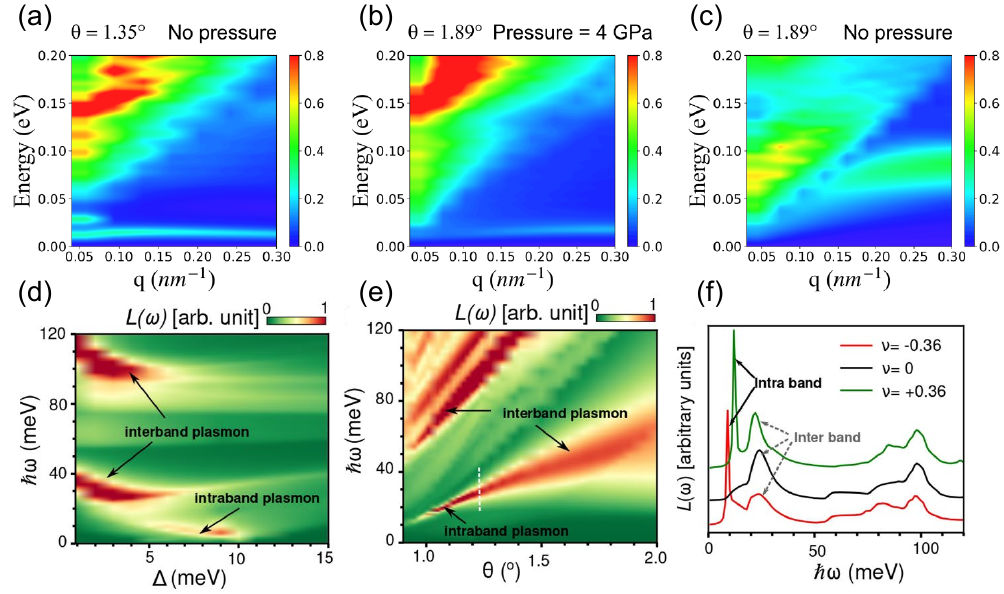}
	\caption{(a) Loss function intensity plots in the magic-angle $1.35^\circ$ twisted trilayer graphene; (b) and (c) pressure engineering plasmon modes in $1.89^\circ$ twisted trilayer graphene. Reprinted with permission from \cite{Wu2021magicangle} Copyright (2021) by the American Physical Society. Twisted double bilayer graphene intraband and interband plasmons as a function of: (d) electric bias, (e) twist angle, and (f) doping level. Reprinted with permission from \cite{Chakraborty2022tunable} Copyright (2022) by the American Physical Society.}
	\label{twmulplas}
\end{figure}

Flat band and Dirac bands are found to coexist in twisted trilayer graphene with mirror symmetry~\cite{hao2021electric,park2021tunable}. This coexistence may allow the plasmons to have different properties from those in TBG. Theoretically, Wu \emph{et al.}~\cite{Wu2021magicangle} numerically investigated plasmons in twisted trilayer graphene with different twist angles and vertical pressures. In particular, for a twist angle of $1.35^{\circ}$, the defined magic angle at which Fermi velocity is zero in this system, a clear quasi-flat plasmon mode emerged below 0.05 eV, as seen in Figure \ref{twmulplas}(a). This plasmon was found to be originated from collective excitations inside the flat bands. As shown in Figure \ref{twmulplas}(c), for large twist angles and no pressure, the quasi-flat plasmon mode had a blue shift to an energy of 0.06 eV indicating the presence of wider bands near charge neutrality. By applying a vertical pressure, the plasmon mode reappeared, indicating an enhancement of the interlayer coupling with pressure. 

On the other hand, numerical studies found that long lived, flat intraband and interband plasmons can exist in twisted double bilayer graphene~\cite{Chakraborty2022tunable}. In particular, it was found that a flat intraband plasmon modes emerged at long momentum because of the influence of higher interband transitions. Furthermore, as shown in Figure \ref{twmulplas}(d)-(f) these plasmon modes were found to be tuned by a vertical electric field, twist angle and doping, respectively. Gapped interband plasmon and intraband plasmon appeared at small and large electric field, respectively, and they persisted over a wide range of twist angles.

\subsection{Plasmons of twisted bilayer transition metal dichalcogenides}
\begin{figure}[h]
	\includegraphics[width=1\textwidth]{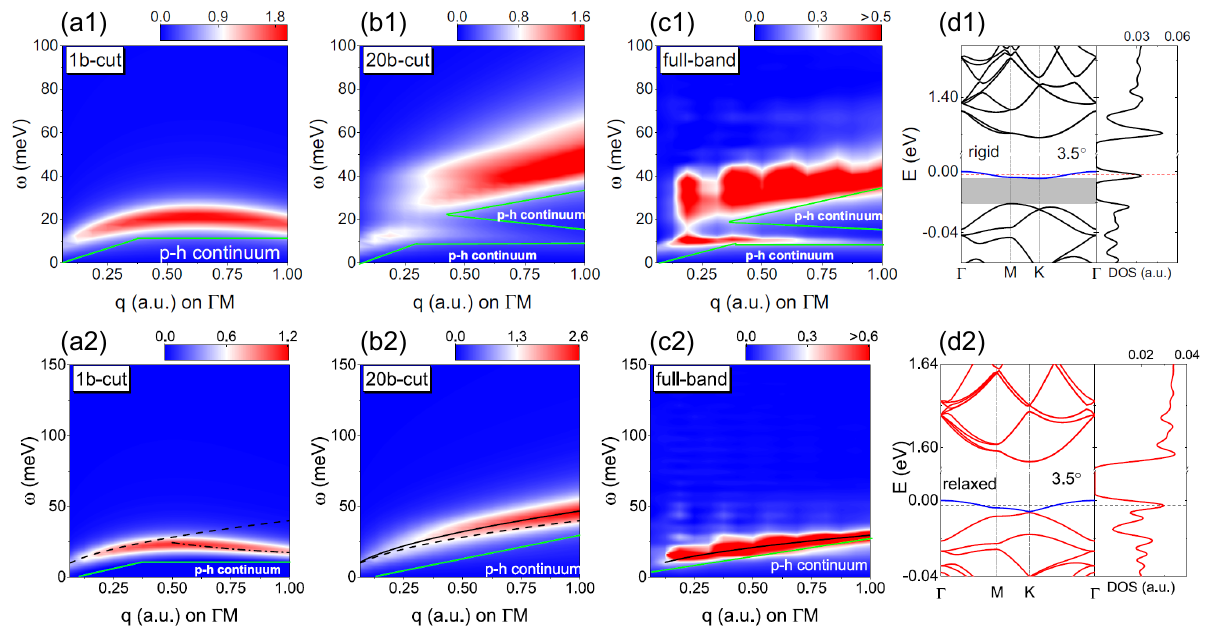}
	\caption{Loss function intensity plots of twisted bilayer MoS$_2$ with (first row) and without (second row) atomic relaxation. (a) Only one band, the flat band included in equation (\ref{lindhard}), while (b) 40 band near the flat band included and (c) full band of tight-binding model with TBPM as equation (\ref{kubo_dyn}). Particle-hole (p-h) continuum region is marked with “p-h continuum” and boundaries with green solid lines. (d) The band structure of $3.5^\circ$ twisted bilayer MoS$_2$ without lattice relaxation (d1) and within (d2). Reprinted with permission from \cite{Kuang2022flatband}  Copyright (2022) by the American Physical Society.}
	\label{tmdplas}
\end{figure}
 In moir\'{e} TMDCs, the existence of flat bands also provides possibility to explore quasi-flat plasmon modes. A recent numerical study~\cite{Kuang2022flatband} suggested that both atomic relaxation and high energy bands have an impact on the low energy flat-band plasmon in twisted bilayer MoS$_2$, shown in Figure~\ref{tmdplas}. In particular, for an unrelaxed system, shown in Figures \ref{tmdplas}(b1) and (c1), a flat intraband and linear interband plasmon modes were found. The distinct results between the different approximations suggested that the interband transitions play an important role in the unrelaxed system. However, the relaxation effects transformed the two plasmon modes to one mode with $\sqrt{q}$ dispersion as seen in Figure \ref{tmdplas}(b1) and (c2). Further analysis concluded that the isolation of the flat band shown in Figure \ref{tmdplas}(d1) was the key to obtain quasi-flat plasmon modes in twisted bilayer TMDCs, and the high-energy interband transitions had impact on plasmons at a large momentum limit \cite{Kuang2022flatband}. In addition to the twist angle effect on plasmons in twisted bilayer MoS$_2$, a recent experiment also showed that film thickness ratio of bilayers could manipulate plasmon topology in twisted WTe$_2$ films\cite{wang2023twisangle}. Single-layer MoS$_2$ could provide multi-component plasmons since it features spin and valley as two extra degree of freedom \cite{Xiao2017multicomponent}, which could be used to engineer the plasmon properties in twisted bilayer TMDCs in future studies.

\section{Relation with other properties}

\subsection{Many-body effects}
\begin{figure}[t]
\includegraphics[width=1\textwidth]{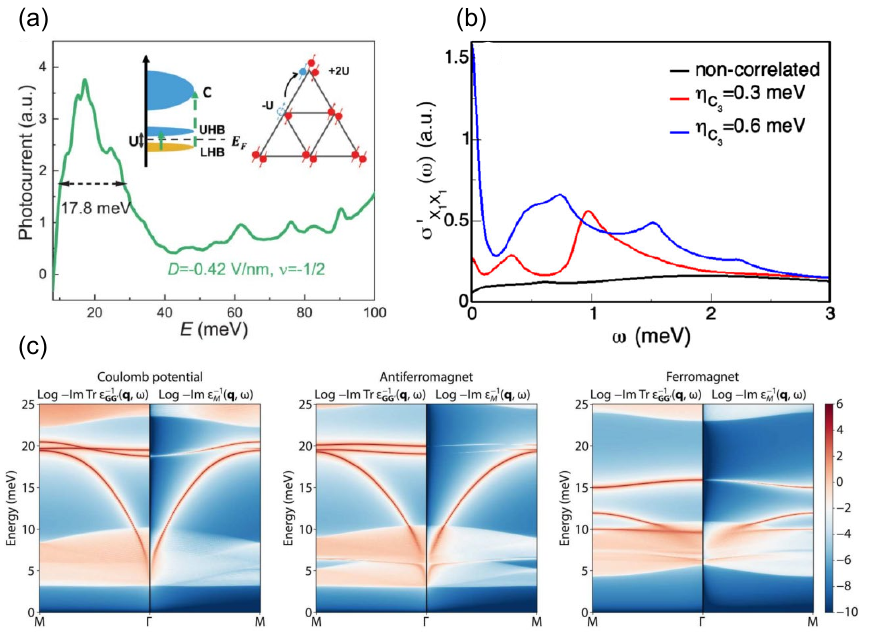}
\caption{Many-body effects. (a) Observed optical response peak in correlated ABC trilayer graphene-hBN moir\'{e} structure \cite{Yang2022spectroscopy}. The photocurrent peak corresponds to the optical transition crossing the Mott gap between lower Hubbard band (LHB) and upper Hubbard band (UHB), as illustrated by the solid arrow in the left inset.The Reprinted with permission from AAAS. (b) Optical conductivity in TBG with correlated effect (nematic order, red and blue lines) and without correlated effect (black line). $\mathrm{\eta_{C_3}}$ is the magnitude of the nematic order. Reprinted from \cite{Calderon2020correlated} CC BY 4.0. (c) Plasmon spectrum of TBG within different many-body effect. Reprinted from \cite{Papaj2022probing} CC BY 4.0.}
\label{optmany}
\end{figure}
Many-body effects can provide a significant effect on optical properties and plasmons of moir\'{e} structures. For example, optical spectroscopy was employed to probe correlations in ABC rhombohedral trilayer graphene with hBN (ABC-hBN) \cite{Yang2022spectroscopy}. In this work, an optical absorption peak emerged at $\sim$ 18 meV, indicating a direct optical excitation across an emerging Mott insulator, as shown in Figure \ref{optmany}(a). A similar optical spectra was observed at different fillings. The optical response was found to be a useful tool to characterize the onsite Coulomb repulsion energy, U, in the corresponding Hubbard model. On the other hand, optical conductivity was also theoretically used to reveal the nature of correlated states in TBG \cite{Calderon2020correlated}. Comparing to the optical excitation that a Drude peak emerged at charge neutrality in non-correlated models, new absorption peaks appeared in the optical spectrum for different values of correlated nematic order parameters $\mathrm{\eta_{C_3}}$, see Figure \ref{optmany}(b). Additional calculations in the same system showed that the optical conductivity can be used to distinguish different symmetry broken states. The plasmons can also be used to probe many body effects as discussed in section 4.1. Recently, Papaj \emph{et al.}\cite{Papaj2022probing} proposed to probe correlated states with plasmons in twisted heterobilayer TMDCs, where a folded plasmon spectrum can be a signature of correlated states, as shown in Figure \ref{optmany}(c). Here, the plasmon spectra has different characteristics depending on the type of correlated effects.

\subsection{Non-linear optical response} 
\begin{figure*}[h]
    \centering
    \includegraphics[scale =0.35]{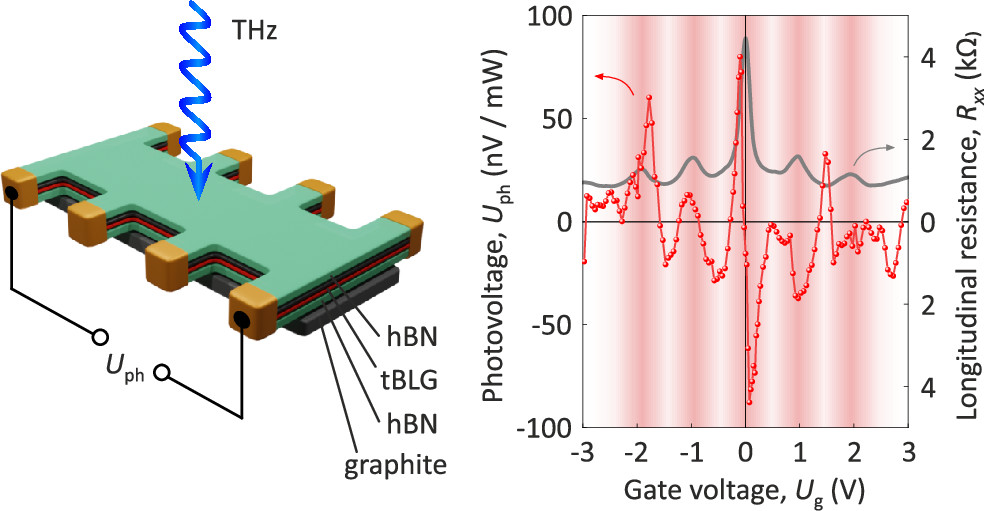}
    \caption{Photogalvanic effect in low angle TBG ($\theta = 0.6^\circ$). (Left side) Schematics of a TBG sample encapsulated by hBN. (Right side) An incident terahertz radiation gives rise to a photovoltage. Reprinted with the permission from~\cite{Otteneder2020Terahertz}. Copyright {2020} American Chemical Society.}
    \label{fig:galvanicTBG}
\end{figure*}
The nonlinear optical response in TBG, also referred to as optotwistronics~\cite{Ikeda2020HighOrder} or twistoptics~\cite{Herzig2020Rise}, has attracted attention only recently. Theoretical works have explored the impact of light on the TBG band structure. In particular, a Floquet band engineering has been investigated by means of tight-binding models~\cite{topp2019Topological,Du2021HighOrder}. Due to the extensive number of sites in the TBG {\mo} unit cell, the continuum model has proven to be highly valuable for studying optically induced flat bands~\cite{Katz2020Optically}, the manipulation of interlayer couplings~\cite{Vogl2020Floquet}, and the formulation of effective Floquet Hamiltonians~\cite{Vogl2020Effective}.

As mentioned in the previous section, the TBG transition energies are notably influenced by the twist angle~\cite{brihuega2012unraveling,Zuber2021NonLinear}. Consequently, for a constant twist angle, the activation or deactivation of one- and two-photon resonances can be achieved by adjusting the incident wavelength. This characteristic allows for a highly adaptable second harmonic generation in TBG~\cite{Yang2020Tunable,Du2020Twisting}. Remarkably, at larger twist angles ($\sim 21.79^\circ$) and in the presence of intense laser fields, TBG has been found to display high-harmonic generation, that by symmetry grounds cannot occur in monolayer or bilayer graphene~\cite{Ikeda2020HighOrder,DiMauro2022Optotwistronics}. The selection rules described in~\cite{Ikeda2020HighOrder} under circularly polarized light stem from the underlying lattice symmetry of TBG at large angles.

On the other hand, at low twist angles ($\sim 1.05^\circ$) there is an interplay between twist, band geometry and optical response~\cite{DiMauro2022Optotwistronics}. There is an emergence of dynamical symmetries coupled with the standard symmetries of the TBG lattice ($\mathcal{C}_{2y}$ and $\mathcal{C}_3$) which are not present at large twist angles ($\sim 21.79^\circ$) . Additional selection rules for the current response are obtained at the low angle regime. In particular, an induced inversion symmetry breaking~\cite{Shimazaki2015Generation} in the bilayer system allows for a non-zero finite Berry curvature which enhances the non-linearity. Interestingly, as described in Ref.~\cite{DiMauro2022Optotwistronics} the high order harmonics were found strongly dependent on the magnitude of the band geometry through the Berry curvature. The photogalvanic effect, which is the lowest order non-linear effect, was predicted~\cite{Gao2020Tunable,Otteneder2020Terahertz} and then experimentally observed in low-angle ($\sim 0.6^\circ$) samples of TBG~\cite{Otteneder2020Terahertz}, see Figure~\ref{fig:galvanicTBG}. At higher orders, a third order optical non-linearity was also reported~\cite{Ha2021Enhanced}, where the non-linear response was found to be considerably modified depending on the rotation angle in TBG. Effects of the band topology~\cite{Hu2023berry}, high harmonic generation~\cite{Molinero2023highhamonic}, correlated insulating states~\cite{Zhang2022correlated} have also been investigated. In addtition to TBG, non-linear optical response can be achieved in large-angle ($\sim 50^\circ$) twisted multilayer WS$_2$\cite{kim2023three}. It is worthy to mention that via a second harmonic generation in heterostructures of graphene-hBN a transtion from a commensurate to a non-commensurate state was detected~\cite{Stepanov2020direct}.

\subsection{Superconductivity}
\begin{figure}[h]
\includegraphics[width=1\textwidth]{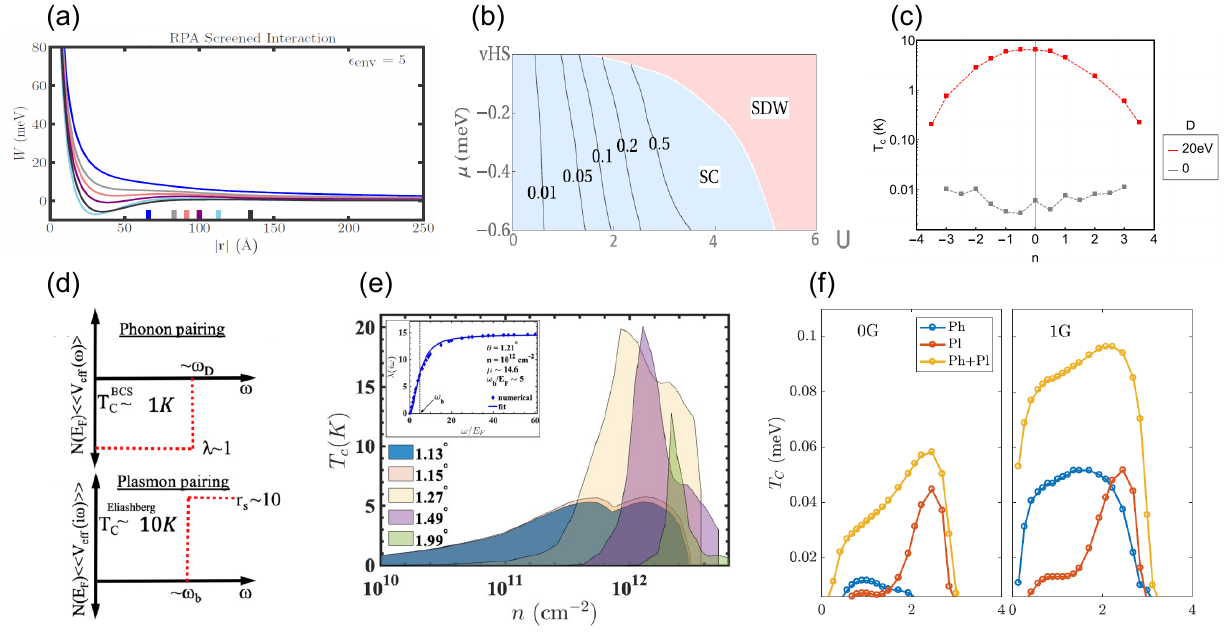}
\caption{(a) Attractive electron-electron interaction in magic-angle twisted bilayer graphene. Reprinted with permission from \cite{Goodwin2019attractive} Copyright (2019) by the
American Physical Society. (b) Purely electron-electron induced superconductor phase in magic-angle twisted bilayer graphene. Reprinted with permission from \cite{Gonazlez2019kohn} Copyright (2019) by the
American Physical Society. (c) Superconductivity critical temperature explained by electron-electron screening potential with (red dots) and without phonon effect (gray dots) Reprinted from \cite{Cea2021coulomb} CC BY-NC-ND 4.0. (d) Plasmon paring driven by dynamic Coulomb interaction comparing to Phonon paring and (e) critical temperature of plasmon-mediated superconductivity in TBG at different angles. Reprinted from \cite{Sharma2020superconductivity} CC BY 4.0. (f) Phonon (Ph) and plasmon (Pl) mediated superconductivity in magic-angle TBG without (0G) and within (1G) local field effect. Reprinted with permission from \cite{Lewandowski2021paring} Copyright (2021) by the
American Physical Society.}
\label{plassuper}
\end{figure}

The origin of superconductivity in {\mo} twisted graphene layers remains a subject of debate~\cite{andrei2021marvels,torma2022superconductivity}. In addition to the proposed phonon-mediated theory~\cite{Wu2018theory,Lian2019phonon}, purely electronic mechanisms such as the Kohn-Luttinger (KL) mechanism and plasmon-mediated superconductivity have also been examined~\cite{Gonazlez2019kohn,Sharma2020superconductivity,Lewandowski2021paring}. This is motivated by the significant influence of electron-electron interactions found in magic-angle TBG~\cite{cao2018unconventional,cao2018correlated}.\ For instance, theoretical studies of the dielectric function within the RPA, cf. equation (\ref{V_scr}), reveal that the screened Coulomb potential calculated near the magic-angle TBG displays attractive regions in real space~\cite{peltonen2018mean,Gonazlez2019kohn,Cea2021coulomb}, see Figure \ref{plassuper}(a), indicating that superconductivity could be induced by pure electron-electron interactions~\cite{Guinea2018,Cea2023Superconductivity} through a KL mechanism\cite{Gonazlez2019kohn,Cea2022Electrostatic}. The RPA calculations with frequency-independent polarization function show that the superconducting instability can appear near VHS and precedes a spin-density-wave instability under KL mechanism, as shown in Figure~\ref{plassuper}(b). Furthermore, the critical temperature has also been predicted based on the KL mechanism by calculating the static screened dielectric function and the gap equation~\cite{Cea2023Superconductivity}. However, it is worth noting that the predicted critical temperature is not as high as what has been observed in experiments, as indicated by the gray dots in Figure~\ref{plassuper}(c). 
This suggests that phonons may also contribute to the enhancement of the superconducting pairing\cite{Cea2021coulomb}.

On the other hand, as depicted in Figure \ref{plassuper}(d), the intrinsic plasmon mode in magic-angle TBG is estimated to lead to an even higher critical temperature (T$_c$) than the effects of phonons under a massless Dirac model. Further numerical calculations reveal that the dynamical Coulomb-driven T$_c$ can reach approximately 15 K and varies with the electron density of TBG near the magic angle, as shown in Figure \ref{plassuper}(e). Further research delves into how the local field effect (LFE) plays a role in cooperative effects between plasmons and phonons on superconductivity\cite{Lewandowski2021paring}. As illustrated in Figure \ref{plassuper}(f), plasmon-mediated superconductivity appears to be insensitive to LFE, which aligns with previous findings indicating that plasmons in TBG are not significantly affected by LFE\cite{stauber2016plas,Lewandowski2019intrinsically}. Recently, extrinsic screening effects on superconductivity were investigated in TBG as well, showing that the critical temperature was unaffected by screening unless the screening layer was lower than three nanometers from the superconductor\cite{peng2023theoretical}. Besides LFE, effect of vertex corrections on plasmon-mediated superconductivity in {\mo} structures could deserve some attentions\cite{martin2020moire}, while vertex corrections could be safely dropped in conventional high-density superconductors according to Migdal's theorem\cite{migdal1958interaction,takada1992insignificance}.

\section{Summary and Outlook} 
This paper reviews recent theoretical and experimental research on optical properties and plasmons in widespread non-twisted and twisted {\mo} structures. 
For non-twisted {\mo} graphene-based structures, the {\mo} potential plays a key role in producing saddle points in the miniband structure of graphene, which gives rise to new optical interband transitions between VHS which are ultimately reflected in the experimental optical conductivity and interband plasmons. In particular, particle-hole asymmetric features in optical conductivity and plasmons emerged due to {\mo} potential breaking the symmetry of electronic wavefunction in graphene-hBN structure. The {\mo}-induced interband plasmons and intrinsic intraband plasmon can coexist in graphene-based non-twisted {\mo} heterostructures.  In twisted {\mo} structures, changing the twist angle results in varying {\mo} lengths, leading to a reshaping of the band structure and alterations in both bandwidth and band velocity. 
In most studies on TBG, the optical absorption peaks and plasmon wavelengths exhibit a red shift as the twist angle decreases. Additionally, interband plasmons, low-damped, and slow plasmon modes have been theoretically explored in flat-band TBG and confirmed by experiments. These findings have spurred further investigations into plasmonics in flat-band twisted trilayer graphene, double bilayer graphene, and twisted TMD systems. Twisting is also recognized as a means to modulate the optical response in other twisted 2D {\mo} systems that also exhibit intriguing optical phenomena.
In flat-band {\mo} structures, electron-electron interactions can also have an impact on the optical response. Some studies have focused on understanding the effects of these interactions through optical dynamics and plasmonics. Additionally, the formation of superconducting electron pairs through plasmon and electron-electron interactions under the random phase approximation provides insights into the mechanisms underlying superconductivity in {\mo} structures.

Prosperous and tremendous theoretical and experimental studies on {\mo} structures are still ongoing to open new avenues for physics and potential applications. Twist-angle induced {\mo} potentials are appearing in other structures, such as {\mo} of {\mo} graphene layers\cite{uri2023superconductivity,devakul2023magic,meng2023commensurate}, TBG-hBN heterostrucuture\cite{Cea2020Band,shi2021moire,long2022atomistic,Long2023Electronic}, TBG-TMDC heterostructure\cite{Lin2022spinorbit}, twisted three-dimensional systems\cite{fujimoto2022perfect,song2021eshelby,mullan2023mixing}, and so on. Beyond aforementioned non-twisted {\mo} heterostructures, studies of MoS$_2$-metal {\mo} systems\cite{reidy2021direct}, and non-twisted TMDC bilayer {\mo} heterostructures such as MoTe$_2$/WSe$_2$ are also arising\cite{Li2021quantum,Li2021continuous}. These {\mo} structures could also be ideal platforms for exploring interesting linear and non-linear optical properties as well as plasmonics. For example, spin-orbit coupling accompanying with the {\mo} potentials could cause more optical transitions in TBG-TMDC systems; the hBN induced band gap could change the plasmon energy in TBG-hBN {\mo} structures. 
In addition, the twist-angle and electron-electron interaction effect can also be important in aligned graphene-based heterostructures in prospective studies. Disorder effects, such as twist-angle and strain effects, have shown an impact on electronic and transport properties in {\mo} structures\cite{uri2020mapping,kazmierczak2021strain,wilson2020disorder,namarvar2020electronic,nakatsuji2022moire}. More studies of disorder effects on optical properties of {\mo} structures are also needed in future.
Last but not least, although the optical and plasmonic applications of {\mo} structures are not the primary focus of this topic review they deserve further investigation and attention.

Finally, by creating {\mo} patterns and further tuning them as needed (e.g., by adjusting the twist angle, combining different materials, and applying artificial structure potentials~\cite{forsythe2018band}), we can obtain control over the manipulation of light in future state-of-the-art technologies. This control may find applications in {\mo}  photonics and {\mo}  optoelectronics, including lasers, detectors, modulators, infrared/terahertz photoresponses, and polarizers. Exploring and gaining a fundamental understanding of how the {\mo} potential influences the optoelectronic properties of these materials is, therefore, crucial for the advancement of the field.\\

~\\
\textbf{Data availability statement}
~\\
Any data that support the findings of this study are included within the article.

\section*{Acknowledgements}
\addcontentsline{toc}{section}{Acknowledgements}
We thank Tommaso Cea, Alejandro Jimeno-Pozo, H\'ector Sainz-Cruz, Yunhai Li, Zewen Wu, Yonggang Li for fruitful discussions. IMDEA Nanociencia acknowledges support from the ``Severo Ochoa" Programme for Centres of Excellence in R\&D (CEX2020-001039-S / AEI / 10.13039/501100011033). P.A.P and F.G. acknowledge funding from the European Commission, within the Graphene Flagship, Core 3, grant number 881603 and from grants NMAT2D (Comunidad de Madrid, Spain), SprQuMat and (MAD2D-CM)-MRR MATERIALES AVANZADOS-IMDEA-NC. S.Y. acknowledges support funding from the National Natural Science Foundation of China (No 12174291) and Natural Science Foundation of Hubei Province, China (Grant No. 2022BAA017). Z.Z. acknowledges support funding from the European Union's Horizon 2020 research and innovation programme under the Marie Skłodowska-Curie grant agreement No 101034431 and from the ``Severo Ochoa" Programme for Centres of Excellence in R\&D (CEX2020-001039-S / AEI / 10.13039/501100011033).
\vspace{2cm}
\bibliographystyle{iopart-num}
\bibliography{reference_review}
	
\end{document}